\documentclass[usenatbib,usedcolumn]{mnras}
\usepackage[british]{babel}             
\usepackage[T1]{fontenc}                
\usepackage{graphicx}                   
\usepackage[usenames,dvipsnames]{color}

\hypersetup{pdfauthor={I. Heywood},
               pdftitle={MIGHTEE: total intensity radio continuum imaging and the COSMOS / XMM-LSS Early Science fields},
               pdfkeywords={techniques: interferometric},
               bookmarksnumbered=true}

\volume{{\rm in press}}

\usepackage{graphicx}	
\usepackage{amsmath}	
\usepackage{amssymb}	
\usepackage{color}

\definecolor{mygray}{gray}{0.6}
\newcommand{\hhh}{$^{\mathrm{h}}$}
\newcommand{\mmm}{$^{\mathrm{m}}$}
\newcommand{\sss}{$^{\mathrm{s}}$}
\newcommand{\ddd}{$^{\mathrm{\circ}}$}
\newcommand{\dmm}{$^{\prime}$}
\newcommand{\dss}{$^{\prime\prime}$}

\title[MIGHTEE continuum processing and Early Science]{MIGHTEE: total intensity radio continuum imaging and the COSMOS / XMM-LSS Early Science fields}

\author[Heywood et al.]
{\parbox{\textwidth}{
\begin{flushleft}
I.~Heywood$^{1,2,3}$\thanks{E-mail: ian.heywood@physics.ox.ac.uk},
M.~J.~Jarvis$^{1,4}$,
C.~L.~Hale$^{5}$,
I.~H.~Whittam$^{1,4}$,
H.~L.~Bester$^{3,2}$,
B.~Hugo$^{3,2	}$,
J.~S.~Kenyon$^{2,3}$,
M.~Prescott$^{4,6}$,
O.~M.~Smirnov$^{2,3}$,
C.~Tasse$^{7,2}$,
J.~M.~Afonso$^{8,9}$,
P.~N.~Best$^{10}$,
J.~D.~Collier$^{11,12,13}$,
R.~P.~Deane$^{14,15,2}$,
B.~S.~Frank$^{3,11,16}$,
M.~J.~Hardcastle$^{17}$,
K.~Knowles$^{2,3}$,
N.~Maddox$^{18}$,
E.~J.~Murphy$^{19}$,
I.~Prandoni$^{20}$,
S.~M.~Randriamampandry$^{21,22,23}$,
M.~G.~Santos$^{4,3}$,
S.~Sekhar$^{11,24,4}$,
F.~Tabatabaei$^{25,26}$,
A.~R.~Taylor$^{11,16,4}$,
K.~Thorat$^{15}$
\\
\end{flushleft}
}
\footnotesize
\\
Author affiliations are listed at the end of the paper.
}

\date{Accepted XXX. Received YYY; in original form ZZZ}

\pubyear{2021}

\begin{document}
\label{firstpage}
\pagerange{\pageref{firstpage}--\pageref{lastpage}}
\maketitle

\begin{abstract}

MIGHTEE is a galaxy evolution survey using simultaneous radio continuum, spectro-polarimetry, and spectral line observations from the South African MeerKAT telescope. When complete, the survey will image $\sim$20 deg$^{2}$ over the COSMOS, E-CDFS, ELAIS-S1, and XMM-LSS extragalactic deep fields with a central frequency of 1284 MHz. These were selected based on the extensive multiwavelength datasets from numerous existing and forthcoming observational campaigns. Here we describe and validate the data processing strategy for the total intensity continuum aspect of MIGHTEE, using a single deep pointing in COSMOS (1.6 deg$^{2}$) and a three-pointing mosaic in XMM-LSS (3.5 deg$^{2}$). The processing includes the correction of direction-dependent effects, and results in thermal noise levels below 2~$\mathrm{\mu}$Jy beam$^{-1}$ in both fields, limited in the central regions by classical confusion at $\sim$8$''$ angular resolution, and meeting the survey specifications. We also produce images at $\sim$5$''$ resolution that are $\sim$3 times shallower. The resulting image products form the basis of the Early Science continuum data release for MIGHTEE. From these images we extract catalogues containing 9,896 and 20,274 radio components in COSMOS and XMM-LSS respectively. We also process a close-packed mosaic of 14 additional pointings in COSMOS and use these in conjunction with the Early Science pointing to investigate methods for primary beam correction of broadband radio images, an analysis that is of relevance to all full-band MeerKAT continuum observations, and wide field interferometric imaging in general. A public release of the MIGHTEE Early Science continuum data products accompanies this article.

\end{abstract}




\section{Introduction}

Radio continuum observations are a uniquely powerful tool in the pursuit of understanding how galaxies form and evolve over cosmic time, one of the key goals of modern astrophysics. Radio emission arises from the cores of active galactic nuclei \citep[AGN;][]{white2015,whittam2016}, and their powerful jet-driven radio lobes \citep{laing2011,fanaroff2021}; star-formation in `regular' galaxies can be detected via generally-fainter synchtrotron emission \citep{condon1992,jarvis2010,delvecchio2021} as well as from thermal emission at higher radio frequencies \citep{murphy2017}; and, on Mpc scales, radio observations reveal diffuse radio haloes that trace the hot gas within galaxy clusters, as well as the shock-driven relic structures that can be found on their peripheries \citep{vanweeren2019}. Advances in instrumentation and data processing methods over the last decade have led to a new generation of large-scale radio surveys that are targeting the emission processes described above to efficiently gather galaxy detections in statistically significant numbers, and in a range of environments out to the highest redshifts.

Telescope time is a finite commodity, and the most efficient way to conduct extragalactic surveys is to adopt a tiered approach whereby areal coverage is traded for observational depth. Examples of on-going surveys that are essentially covering the entire sky visible to their respective observatories are the Australian Square Kilometre Array's \citep[ASKAP;][]{hotan2021} Rapid ASKAP Continuum Survey \citep[RACS;][]{mcconnell2020,hale2021} and forthcoming Evolutionary Map of the Universe \citep[EMU;][]{norris2015} survey, the Very Large Array Sky Survey \citep[VLASS;][]{lacy2020}, and at lower radio frequencies the Galactic and Extragalactic All-sky Murchison Widefield Array survey \citep[GLEAM;][]{hurleywalker2017} and the LOFAR Two-metre Sky Survey \citep[LoTSS;][]{shimwell2017} being conducted on the Murchison Widefield Array \citep[MWA;][]{lonsdale2009} and Low Frequency Array \citep[LOFAR;][]{vanhaarlem2013} respectively. Complementary to these are intermediate tier surveys that have higher sensitivities over areas of $\sim$100s--1000s of square degrees, such as the VLA Stripe 82 surveys \citep{heywood2016,mooley2016}, the 325~MHz observations of the Galaxy and Mass Assembly (GAMA) fields with the Giant Metrewave Radio Telescope \citep{mauch2013}, and the imaging surveys with the Westerbork Synthesis Radio Telescope's Aperture Tile In Focus (APERTIF) upgrade \citep{vanCappellen2021}. Finally, over smaller areas, typically 1--10 deg$^{2}$, we find the deepest radio surveys reaching $\mathrm{\mu}$Jy sensitivities, well into the regime where regular star-forming galaxies are dominating the radio source counts \citep[e.g.][]{prandoni2018, matthews2021}, for example the deep VLA observations of the COSMOS field \citep{smolcic2017,vandervlugt2021}, the VLA Hubble Frontier Fields survey \citep{heywood2021}, the e-MERLIN e-MERGE project \citep{muxlow2020}, the LoTSS deep fields \citep{tasse2020}, and the `DEEP2' observations \citep{mauch2020} with the South African MeerKAT telescope \citep{jonas2016}.

 MeerKAT consists of 64 $\times$ 13.5~m dishes with offset Gregorian optics, providing an unblocked aperture. It is equipped with three receiver bands; UHF (544 -- 1088 MHz), L-band (856 -- 1712 MHz), and S-band (1750 -- 3500  MHz). Three-quarters of the collecting area is within a dense, 1~km diameter core region, and the remaining dishes are situated around the core, providing a maximum baseline of 8~km. The large number of baselines, wide field of view (1 deg at L-band), and low ($\sim$20 K) system temperature all conspire to make MeerKAT an exceptionally fast and capable synthesis imaging telescope. The correlator can also deliver up to 32,768 frequency channels, delivering excellent spectroscopic imaging capabilities.

Also coming in at the deep end is MIGHTEE \citep[MeerKAT International Gigahertz Tiered Extragalactic Explorations;][]{jarvis2016}. MIGHTEE is one of MeerKAT's flagship Large Survey Projects, a galaxy evolution survey that is using simultaneous continuum, polarimetry (Sekhar et al., \emph{in prep.}) and spectral line \citep{maddox2021} measurements to investigate the formation and evolution of galaxies over cosmic time. It will use $\sim$1000 h of observations with MeerKAT's L-band receivers, with the goal of imaging 20 square degrees over four extragalactic deep fields, namely COSMOS, the Extended Chandra Deep Field South (E-CDFS), the southermost field of the European Large Area ISO Survey (ELAIS-S1), and the XMM-\emph{Newton} Large Scale Structure field (XMM-LSS). The deep multiwavelength data in these fields is essential for obtaining redshifts for the radio sources, and disentangling the relative contributions of star formation and black-hole accretion to the total radio emission \citep{white2017}. The design of the MIGHTEE survey is such that the total intensity continuum images reach the classical confusion limit of MeerKAT with RMS fluctuations of approximately 2$\mathrm{\mu}$Jy beam$^{-1}$. The field selection and the large investment of telescope time from the observatory provide MIGHTEE with an unrivalled combination of depth, area, and corresponding multiwavelength data, pushing both the survey parameters and the resulting science beyond the state of the art.

This article presents a description (Section \ref{sec:obs_and_proc}) and validation (Section \ref{sec:discussion}) of the continuum data processing strategy for the MIGHTEE survey. The wide field of view and extreme instantaneous sensitivity of MeerKAT means that direction dependent calibration methods are necessary to reach the requirements of the survey's design. We use this processing strategy to provide the `Early Science' continuum data products for the MIGHTEE survey (Section \ref{sec:results}), namely a single pointing in the COSMOS field, and a three-pointing mosaic in the XMM-LSS field, from which we extract component catalogues (Section \ref{sec:catalogues}). 

In designing and validating our data processing strategy, we conducted an investigation of some different primary beam correction methods for broadband radio continuum imaging, and we include the results of this investigation here, as it is more broadly relevant for widefield, full-band imaging with MeerKAT, and modern radio interferometers in general. We provide a public data release\footnote{\url{https://doi.org/10.48479/emmd-kf31}} of the catalogue and image products with this article. In the interests of reproducibility, and to provide even more detail on our processing methods for the truly curious, the scripts that were written to process the MIGHTEE data are made available online\footnote{v0.1; \url{https://github.com/IanHeywood/oxkat}} \citep{oxkat2020}. These have already proven to be suitable for general, semi-automatic processing of MeerKAT continuum observations, and continue to be developed.

\section{Observations and data processing}
\label{sec:obs_and_proc}

In addition to providing an initial release of continuum data to the community, the Early Science phase of MIGHTEE has also been used to develop and refine the data processing methods for the full-scale survey. This section provides a summary of the Early Science observations, and a detailed description of the full-band, Stokes I continuum imaging strategy. Numerous software packages are used, as cited in the sections that follow. 

\subsection{MeerKAT observations}
\label{sec:observations}

The properties of the MeerKAT observations that make up the MIGHTEE continuum Early Science products are listed in Table \ref{tab:observations}. In total there are 25 h of observations in COSMOS for an on-source time of 17.45 h. For the three pointings in XMM-LSS there are 16.05, 16.12 and 16.03 h for XMM-LSS\_12, XMM-LSS\_13 and XMM-LSS\_14, with 12.41, 12.45 and 12.43 h of on-source time respectively. The observational setup depends slightly on the vintage of the observing block. Typically, primary calibrators were visited for 5-10 minutes, at least twice per block, and secondary calibrators were visited for 2-3 minutes following every 20-30 minute target scan. As part of the verification of the Early Science data (Section \ref{sec:discussion}) we also image and make use of all of the pointings available in COSMOS at the time of writing, although these image products are not part of the Early Science data release. These additional observations are summarised in Table \ref{tab:extra_observations}. For all observations used in this paper the correlator integration time was 8~s per visibility point.

\begin{table*}
\begin{minipage}{176mm}
\centering
\caption{Properties of the MeerKAT observations of the COSMOS and XMM-LSS fields that form the MIGHTEE Early Science data.}
\begin{tabular}{lllllllllll} \hline
Date             & Block ID         & Field  & RA          & Dec        & Track   & On-source & N$_{\mathrm{chan}}$ & N$_{\mathrm{ant}}$ & Primary    & Secondary  \\ 
(UT, J2000)      &                  &        &             &            & (h)     & (h)       &                     &                    & calibrator & calibrator \\ \hline
2018-04-19 & 1524147354 & COSMOS & 10\hhh00\mmm28\fs6s & +02\ddd12\dmm21\dss & 8.65    & 6.1       & 4096  & 64 & J0408-6545 & 3C237\\
2018-05-06 & 1525613583 & COSMOS & 10\hhh00\mmm28\fs6s & +02\ddd12\dmm21\dss & 8.39    & 5.1       & 4096  & 62 & J0408-6545 & 3C237\\
2020-04-26 & 1587911796 & COSMOS & 10\hhh00\mmm28\fs6s & +02\ddd12\dmm21\dss & 7.98    & 6.25      & 32768 & 59 & J0408-6545 & 3C237\\
& & & & & & & & & & \\
2018-10-06 & 1538856059 & XMMLSS\_12 & 02\hhh17\mmm51\sss & -04\ddd49\dmm59\dss & 8.02   & 6.22      & 4096 & 59 & J1939-6342 & J0201-1132 \\
2018-10-07 & 1538942495 & XMMLSS\_13 & 02\hhh20\mmm42\sss & -04\ddd49\dmm59\dss & 8.07   & 6.22      & 4096 & 59 & J1939-6342 & J0201-1132 \\
2018-10-08 & 1539028868 & XMMLSS\_14 & 02\hhh23\mmm22\sss & -04\ddd49\dmm59\dss & 8.03   & 6.19      & 4096 & 60 & J1939-6342 & J0201-1132 \\
2018-10-11 & 1539286252 & XMMLSS\_12 & 02\hhh17\mmm51\sss & -04\ddd49\dmm59\dss & 8.05   & 6.23      & 4096 & 63 & J1939-6342 & J0201-1132 \\
2018-10-12 & 1539372679 & XMMLSS\_13 & 02\hhh20\mmm42\sss & -04\ddd49\dmm59\dss & 8.03   & 5.92      & 4096 & 62 & J1939-6342 & J0201-1132 \\
2018-10-13 & 1539460932 & XMMLSS\_14 & 02\hhh23\mmm22\sss & -04\ddd49\dmm59\dss & 8.0    & 6.24      & 4096 & 62 & J1939-6342 & J0201-1132 \\ \hline
\end{tabular}
\label{tab:observations}
\end{minipage}
\end{table*}

\subsection{Flagging and reference calibration (1GC)}
\label{sec:processing}

Each individual MeerKAT observation, as itemised in Tables \ref{tab:observations} and \ref{tab:extra_observations}, is a self-contained block containing scans of the target as well as scans of appropriate primary and secondary calibrator sources\footnote{Polarisation calibrators are also included, however we do not make use of them for the total intensity continuum processing.}. These calibrator sources are used to derive corrections for instrumental and propagation effects that are then applied to the target data, a process known as first-generation calibration, or 1GC. 

For each observation we used the KAT Data Access Library\footnote{\url{https://github.com/ska-sa/katdal}} to convert the visibility data into Measurement Set format \citep{kemball2000}. Flags generated by the telescope's control and monitoring system were applied, and frequency averaging was also performed at this point to reduce the number of channels (natively either 4096 or 32768) to 1024. Basic flagging commands were applied to all fields using {\sc casa} \citep{mcmullin2007}. The bandpass edges and the Galactic neutral hydrogen line were flagged for all baselines. Frequency ranges containing persistent radio frequency interference (RFI) were flagged on spacings shorter than 600 m. The auto-flagging algorithms {\sc tfcrop} and {\sc rflag} were used on the calibrator fields with their default settings.

The primary calibrator was used to derive delay and bandpass solutions (both per-scan), and frequency-independent complex gain corrections amplitude and phase terms (per integration time) using the {\sc gaincal} and {\sc bandpass} tasks in {\sc casa}. We used the \citet{reynolds1994} model (see also \citealt{heywood2020b}) to predict the spectral behaviour of the standard calibrator PKS B1934$-$638. In the case of the primary calibrator being PKS 0408$-$65, we assume a point source model with a flux density of 17.066 Jy beam$^{-1}$, and a spectral index\footnote{Note that throughout this paper we adopt the convention that the flux density $S$ is related to frequency $\nu$ via the spectral index parameter $\alpha$: $S~\propto~\nu^{\alpha}$.} of $-$1.179, the reference frequency being 1284 MHz. The calibration is iterative, with rounds of autoflagging on residual (corrected $-$ model) visibilities taking place before the next iteration. 

The gain solutions derived from the primary were applied to the secondary calibrator. We derived complex gain corrections (one solution per scan) from the scans of the secondary calibrator in eight spectral bins. These gains were scaled according to the gain amplitudes derived from the primary calibrator, using the {\sc casa fluxscale} task, allowing us to determine a polynomial model of the intrinsic spectral shape of the secondary calibrator. We then derived per-scan complex gain corrections  from the observations of the secondary using this intrinsic model with the {\sc gaincal} task. This process compensates for the effects of the large fractional bandwidth of MeerKAT. If the secondary calibrator deviates significantly from being spectrally flat, then the flux scale may be biased if this is not taken into account.

The final reference calibration step was to apply the gain solutions to the target data, which were then split out into a separate Measurement Set and flagged using the {\sc tricolour}\footnote{\url{https://github.com/ska-sa/tricolour}} package. Following the removal of the low-gain bandpass edges there is approximately 800 MHz of usable bandwidth. A further loss of about 50 per cent of the data within this region is typical following auto-flagging. The RFI occupancy is strongly dependent on baseline length, and most of this loss occurs on spacings shorter than 1 km in the core of MeerKAT \citep[see also][]{mauch2020}.

\subsection{Direction-independent self-calibration (2GC)}
\label{sec:di_selfcal}

The use of the target data themselves to further refine the antenna-based gain corrections is known as self-calibration, or second-generation calibration (2GC). Multi-frequency synthesis (MFS) images of the target data were made using {\sc wsclean} \citep{offringa2014}. The full-band data were imaged without a cleaning mask, with deconvolution terminating after 100,000 clean components or when the peak residual reaches 20 $\mathrm{\mu}$Jy beam$^{-1}$, whichever occurs first. A clean mask was derived from the resulting image, excluding regions below a local threshold of 6$\sigma$, where $\sigma$ is an estimate of the local pixel standard deviation. We estimate $\sigma$ as a function of position following the method of \citet{tasse2020}. Consider a set of $n$ pixel brightness measurements $X$, drawn from an image with pixels that follow a normal distribution with a mean of 0 and a standard deviation of 1. The cumulative distribution of $Y$~=~$\mathrm{min}\{X\}$ is
\begin{equation}
F = 1 - \left[\frac{1}{2}\left(1 - \mathrm{erf}\left(\frac{y}{\sqrt{2}} \right) \right) \right]^{n}
\end{equation}
where erf is the Gaussian error function. This allows a factor to be derived that converts a measurement of $\mathrm{min}\{X\}$ to $\sigma$ for a given value of $n$, by finding the point where $F(y_{\sigma})~=~\frac{1}{2}$. The number of measurements $n$ is governed by the sliding box size within which $\mathrm{min}\{X\}$ is evaluated for each pixel in the image. The MIGHTEE fields are dominated by compact sources, and we find that a value $\sqrt{n}~=~50$ gives high completeness, returning large numbers of true sources across the full field whilst excluding positive artefacts surrounding brighter sources. Minimum statistics are a good estimator of local noise since thermal noise should be symmetric about zero, and the corrupted PSF that dominates regions blighted by calibration artefacts also has correspondingly lower negatives. The filter is, however, not sensitive to true source confusion which manifests itself as a positive tail on a histogram of pixels drawn from a total intensity radio image. Following the creation of the deconvolution mask the original blind-clean image is discarded and the data are re-imaged using this mask. 

All images extend into the sidelobes of the primary beam in order to deconvolve and model the many sources which are readily detected by MeerKAT in those regions. The total image extent is 10240~$\times$~10240 pixels, each pixel spanning 1$''$.1 for a total image extent of 3.13~$\times$~3.13 deg$^{2}$.

The multi-frequency clean components from the masked image were used to predict a visibility model in eight spectral bins, and the data were self-calibrated using the {\sc casa} {\sc gaincal} task.\footnote{Note that for full-survey MIGHTEE continuum data this step has been replaced by a phase and delay self-calibration operation using {\sc cubical} \citep{kenyon2018}, which has been demonstrated to yield improved results (as implemented in {\sc oxkat v0.2}). The latest version is recommended always.} Frequency-independent phase corrections were derived for every 64 seconds of data, and an amplitude and phase correction was derived for every target scan, with the solutions for the former applied while solving for the latter. The self-calibrated data were then re-imaged using {\sc wsclean}, and the cleaning mask was refined based on the self-calibrated image, and a lower local noise threshold of 5.5$\sigma$.

\subsection{Direction-dependent self-calibration (3GC)}
\label{sec:dd_selfcal}

The dynamic range of the MIGHTEE images is limited by the presence of direction dependent effects (DDEs). At L-band these are principally caused by the time\footnote{Although the MeerKAT primary beam pattern has relatively low sidelobes, the main lobe is not circularly symmetric. The telescope mount causes this pattern to rotate on the sky as an observation progresses. The main lobe variation in the beam holography corresponds to a $\sim$5\% amplitude variation at the nominal half-power point over the course of a full track.}, frequency, and direction dependent variations in the antenna primary beam pattern, coupled with pointing errors. DDEs manifest themselves in the radio images as error patterns resembling a corrupted point spread function (PSF) around off-axis sources of moderate brightness (e.g.~stronger than a few tens of mJy). 

Images from all modern radio telescopes with high sensitivities and broad bandwidths tend to be DDE-limited at some level, and as such many techniques have emerged for dealing with these effects, the so-called third-generation [of] calibration, or 3GC. Visibility-domain treatments include the use of differential gains \citep{smirnov2011b,smirnov2011c}, which is a multi-directional successor to single-source peeling. Approaches that are based on image plane operations include A-Projection, which corrects for DDEs using convolution kernels applied during gridding of the visibilities \citep{bhatnagar2013}, and faceting schemes that apply corrections assuming that DDEs are piecewise constant on a per-facet basis when producing a wide-field image \citep{tasse2018}. For the MIGHTEE continuum processing we primarily adopt the latter image plane facet based approach (Section \ref{sec:ddfacet}), with visibility domain treatments for certain pointings that require it (Section \ref{sec:peeling}).

\subsubsection{Peeling a problem source}
\label{sec:peeling}

Pointings that have a single, dominant, problematic source (typically an off-axis source with an apparent flux density upwards of $\sim$100 mJy beam$^{-1}$) can be subjected to an additional processing step. This involves modelling the source and its associated DDE and subtracting it from the visibility database entirely, a process more generally known as peeling. This is achieved by imaging the data with {\sc wsclean}, with masked deconvolution over a high number of sub-bands (32 by default), and with lower \citet{briggs1995} robust weighting ($-$0.6 by default) in order to obtain a model with high angular and spectral resolution. Model visibilities are then predicted into two separate columns in the Measurement Set, one of which contains only the problem source, and one containing the rest of the sky model. The {\sc cubical} package \citep{kenyon2018} is then used to solve for a direction-independent gain term (\textit{\textbf{G}}, with a default time / frequency interval of 2.4~min / 256 channels) using the complete sky model, whilst simultaneously solving for an additional complex gain term (\textit{\textbf{dE}}, with a default time / frequency interval of 9.6~min / 64 channels) against a model that contains only the problem source. The model of the problem source is then subtracted from the visibilities with \textit{\textbf{dE}} applied, and the residual visibilities containing the rest of the sky are corrected with \textit{\textbf{G}}, ready for subsequent imaging. 

An illustrative Measurement Equation \citep{smirnov2011a} that describes the problem of direction-dependent calibration can be written as
\begin{equation}
\label{eq:me}
V_{pq} = G_{p} \left( \sum_{s} dE_{sp} X_{spq} dE_{sq}^{*} \right) G_{q}^{*}
\end{equation}
where $V$ is the visibility matrix, $p$ and $q$ represent antenna indices, and $X$ is the coherency matrix for source (or direction) $s$. When subtracting a single dominant source $s~=~\{0,1\}$, and the \textit{\textbf{dE}} terms are fixed to unity for $s~=~0$. Direction 0 is represented in this case by the full clean component model covering the entire field of view down to the cleaning threshold, minus the dominant source.

The legacy approach to peeling involves phase-rotation of the visibilities to the direction of the problem source, self-calibrating on that source using a direction-indepenent solver such as the {\sc casa gaincal} task, subtracting the model perturbed by the best-fitting gain solutions, and then undoing the phase-rotation and gain corrections to form the residual visibilities. 

The simultaneous solving of \textit{\textbf{G}} and \textit{\textbf{dE}} offered by {\sc cubical} \citep[and first introduced by {\sc meqtrees};][]{noordam2010} is facilitated by the implementation of an explicit Measurement Equation such as the one above. This has two main advantages over the legacy approach, namely: (i) the inclusion of global \textit{\textbf{G}} solution means that the residual visibilities are less prone to biases such as ghost sources \cite[e.g.][]{grobler2014} and flux suppression  \citep[e.g][]{sardarabadi2019}; (ii) the process is fully general and can subtract multiple problematic sources simultaneously rather than iteratively.

For the Early Science data the peeling step was applied to only two observing blocks, which were those from the XMMLSS\_13 pointing, in order to suppress the effects of the strong compact double source at J2000 02\hhh21\mmm43\fs14 $-$04\ddd13\dmm46\farcs6s. However for the full survey this process will be routine for many of the pointings in E-CDFS and ELAIS-S1, both of which are known to contain a dominant confusing source.

\subsubsection{Facet-based DDE corrections}
\label{sec:ddfacet}

Direction-dependent corrections were made by imaging the visibilities, either those produced following the 2GC stage (Section \ref{sec:di_selfcal}), or those produced by the peeling step (Section \ref{sec:peeling}), with {\sc ddfacet} \citep{tasse2018}. Multiple Measurement Sets containing data from a common pointing centre are brought together at this stage.

The refined cleaning mask produced following self-calibration was used. \citet{hogbom1974} CLEAN was used for decovolution in 8 frequency sub-bands, with the termination thresholds set to 100,000 iterations or a residual peak of 3$\mathrm{\mu}$Jy beam$^{-1}$, ensuring a deep clean is performed within the mask, and a deep sky model is produced. The resulting model was manually partitioned into typically $\sim$10 directions, the exact number depending on the location of off-axis problem sources in the field being imaged, as well as the need to retain suitable flux in the sky model per direction. The {\sc killms} package \citep[e.g][]{smirnov2015} was then used to solve for a complex gain correction for each direction with a time / frequency interval of 5 min / 128 channels. 

Another run of {\sc ddfacet} reimaged the data, applying these directional corrections. For all imaging steps prior to this point (except in the case where a strong source is peeled using a higher resolution model) we use a Briggs' robustness parameter value of $-$0.3. However for the final images we run {\sc ddfacet} twice, with robust values of 0.0 to generate an image optimised for sensitivity, and $-$1.2 to produce a shallower image with higher angular resolution to aid with the multiwavelength cross-matching process.

\begin{figure}
 \includegraphics[width=\columnwidth]{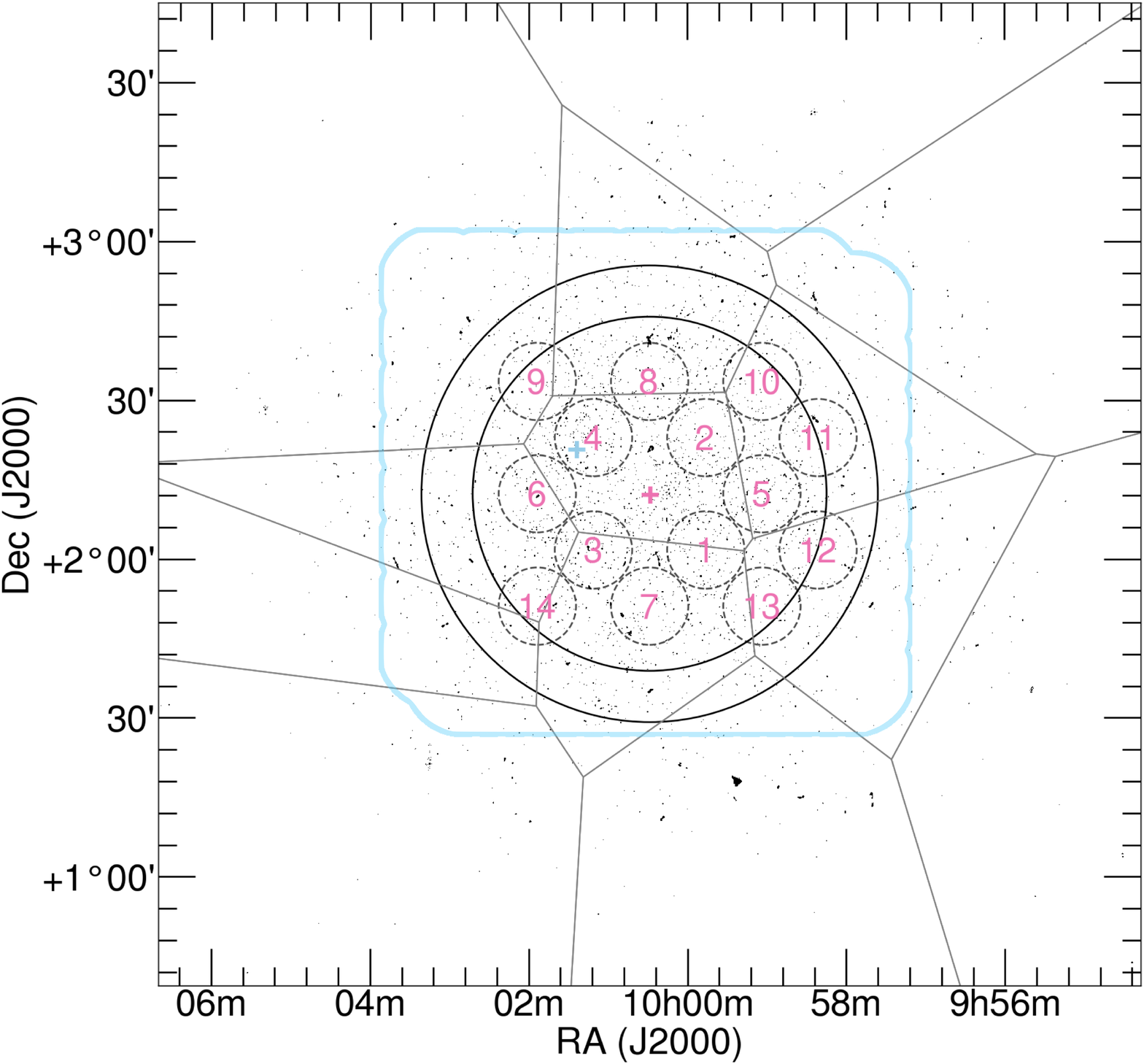}
 \caption{A schematic diagram of the COSMOS field. This shows the full extent of the images formed during the processing of the MIGHTEE continuum data, and the grey polygons show the tesselation pattern over which directional gain corrections are applied in the case of the COSMOS pointing (see Section \ref{sec:obs_and_proc}). The imaged area is significantly larger than the main lobe of the MeerKAT primary beam, the 50\% (0.97 deg$^{2}$) and 30\% (1.62 deg$^{2}$) levels of which are shown by the concentric circles. The central `+' symbol marks the MIGHTEE Early Science COSMOS pointing centre. The numbers mark the locations of the additional pointings used in the analysis in Section \ref{sec:primarybeams}, as summarised in Appendix \ref{sec:cosmosappendix}. The surrounding circles show the range of the nominal 97\% cut of the primary beam pattern for each pointing. The additional `+' marker close to pointing 4 is the pointing centre of the CHILES survey \citep{fernandez2013}, and the blue box shows the coverage of the VLA-COSMOS 3~GHz Large Project \citep{smolcic2017}.}
 \label{fig:schematic}
\end{figure}

An example of the directional partitioning can be seen in Figure \ref{fig:schematic}, with the thirteen regions (tesselations) used in the case of the Early Science COSMOS field (the position of which is shown by the central `+' marker') delineated by the grey convex polygons, representing the patches of sky over which a unique {\sc killms} solution is applied. Figure \ref{fig:schematic} also shows the full area that is imaged, with the inner and outer concentric circles showing the distance from the phase centre at which the primary beam gain has nominally dropped to values of 0.5 and 0.3 respectively. The background image of Figure \ref{fig:schematic} shows a saturated version of the full COSMOS image, and many sources outside the 30\% primary beam level are visible. Please refer to the caption of Figure \ref{fig:schematic} and Section \ref{sec:primarybeams} for further information.

\subsection{Restoring beams, primary beams, and mosaicking}
\label{sec:pbcor}

We perform a post-processing step on the image products in order to impart uniform angular resolution to the mosaicked continuum products, within each of the four MIGHTEE Early Science fields. The clean component model produced by {\sc ddfacet} is convolved with a circular Gaussian, the size of which is selected to slightly exceed the size of the largest major axis (for each of the two weighting settings) from the list of fitted restoring beams for the pointings in and given field. In the case of the Early Science data this was only applicable to XMM-LSS, however the COSMOS field was processed in the same way for consistency\footnote{This process was also used for the full COSMOS mosaic (Section \ref{sec:primarybeams} and Appendix \ref{sec:cosmosappendix}) to give the constituent images matched resolution.}. Following the convolution of the clean component model, we convolved the residual image with a homogenization kernel computed using the {\sc pypher} package \citep{boucaud16}. The goal here is to bring the residual image to a resolution that closely matches that of the restored model, assuming that the fitted restoring beam well-approximates the shape of the main lobe of the synthesised beam (the true PSF). Following these two convolution operations, the model and residual images were summed. The differing of the synthesised beams from pointing to pointing is much less of an issue for MIGHTEE than for larger-area snapshot surveys, as the long tracks and relatively compact mosaics do not give rise to significant variations. However this process ensures that the mosaics can be delivered with restoring beam information in the header that is consistent across each of the constituent fields in a mosaic, which is beneficial for photometric accuracy (see Section \ref{sec:photometry}), as well as image-plane stacking experiments and $P(D)$ measurements \citep{scheuer1957,condon2012}.

The convolved image products for each pointing were corrected for primary beam attenuation by dividing them by a model of the Stokes I primary beam pattern, evaluated at 1284~MHz using the {\sc eidos} \citep{asad2021} package. The main lobe of MeerKAT's primary beam is not circularly symmetric, resulting in a beam gain that varies with azimuthal angle by $\sim$5\% at the nominal half-power point. We smear this variation out by azimuthally-averaging the beam model prior to the division.

Finally, the image products at both resolutions were mosaicked together using the {\sc montage}\footnote{\url{http://montage.ipac.caltech.edu/}} software, using the usual variance-weighted linear combination of pointings, and blanking pixels in each of the constituent images beyond the distance from the phase centre where the beam gain nominally drops below 0.3.

\section{Early science continuum data products}
\label{sec:results}

The observations listed in Table \ref{tab:observations} were processed using the methods described in Section \ref{sec:obs_and_proc}, and the resulting images presented here form the basis of the MIGHTEE Early Science continuum data. The subsections that follow describe the final total intensity images, as well as additional derived image and catalogue products. 

\subsection{Total intensity images}
\label{sec:images}

\begin{figure*}
\centering
\includegraphics[width=6.8in]{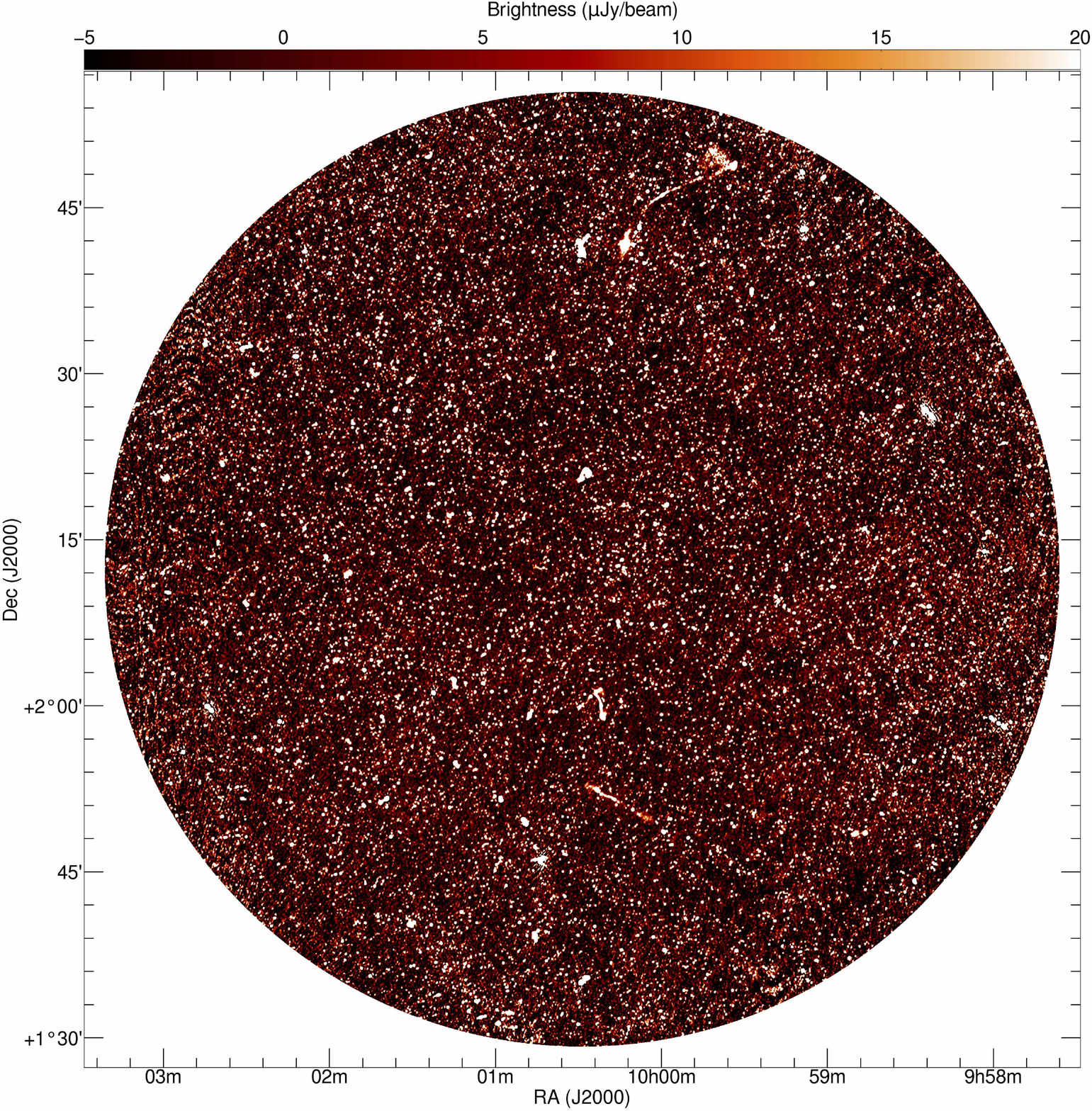}
\caption{The MIGHTEE COSMOS Early Science image. The angular resolution is 8$''$.6, and while the thermal noise in the data is 1.7 $\mu$Jy beam$^{-1}$ this map is limited by classical confusion at approximately 4.5 $\mu$Jy beam$^{-1}$. The image covers 1.6 deg$^{2}$ and contains almost 10,000 radio components with peak brightnesses exceeding 5$\sigma_{\mathrm{local}}$ (see Section \ref{sec:catalogues} for details). The pair of giant radio galaxies reported by \citet{delhaize2021} are visible towards the top and in the lower third of the image.}
\label{fig:cosmos}
\end{figure*}

\begin{figure*}
\centering
\includegraphics[width=6.8in]{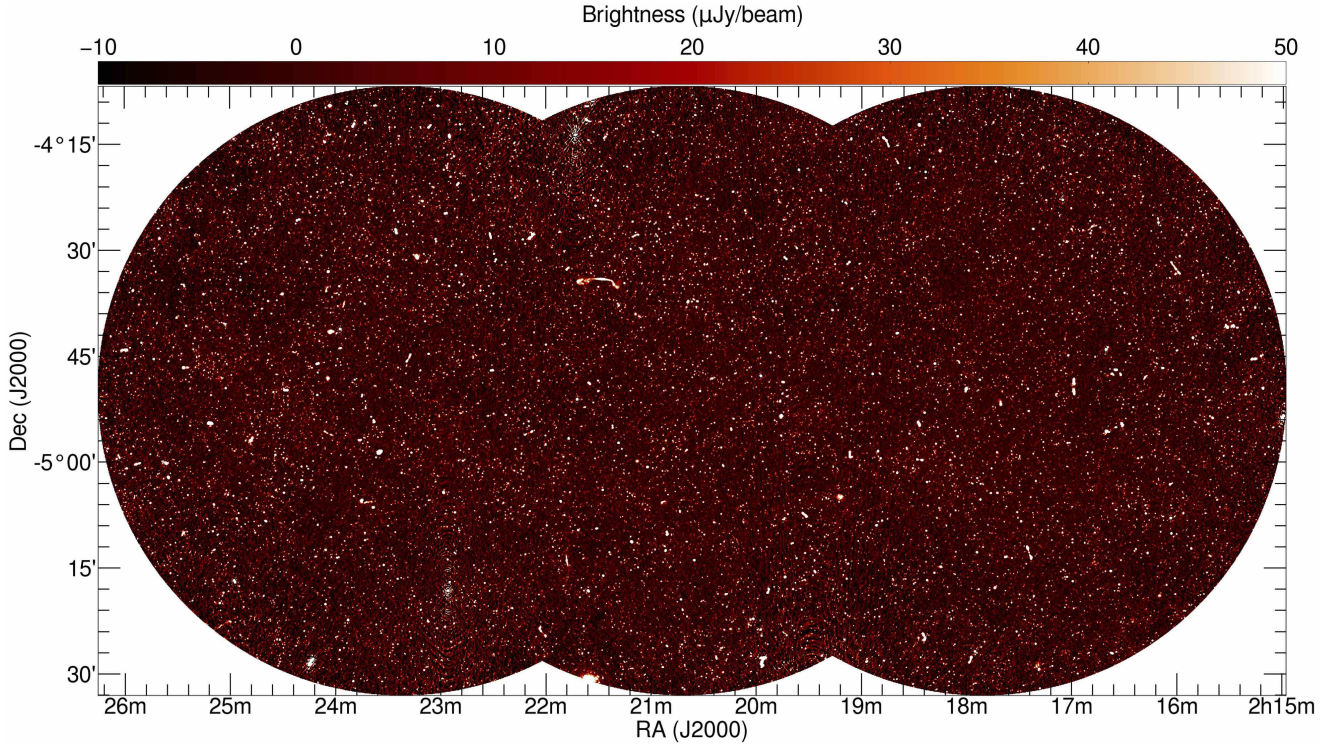}
\caption{The three-pointing MIGHTEE Early Science mosaic covering 3.5 deg$^{2}$ of the XMM-LSS field, with an angular resolution of 8$''$.2.}
\label{fig:xmmlss}
\end{figure*}

The total intensity MIGHTEE Early Science images for COSMOS and XMM-LSS are shown in Figures \ref{fig:cosmos} and \ref{fig:xmmlss} respectively. These images are both the low resolution / high sensitivity variants, imaged with a robust weighting value of 0.0, with angular resolutions of 8$''$.6 and 8$''$.2 for COSMOS and XMM-LSS respectively. 

We measure the thermal noise in these images by taking the RMS of the pixel values in clean, source-free regions away from the main lobe of the primary beam in the images prior to primary beam correction. In the robust 0.0 COSMOS image, the thermal noise is 1.7 $\mathrm{\mu}$Jy~beam$^{-1}$. The deepest combined part of the robust 0.0 XMM-LSS mosaic reaches 1.5 $\mathrm{\mu}$Jy beam$^{-1}$. Both of these values exceed the design goals of the MIGHTEE survey, which was to reach a depth of 2 $\mu$Jy beam$^{-1}$, based on simulations of the angular resolution of MeerKAT, its sensitivity, and the expected classical confusion limit. 

The central regions of the robust 0.0 images are indeed limited by classical confusion rather than thermal noise. Classical confusion imposes a fundamental limit to the depth of an astronomical image, occuring when the surface density of discrete sources increases to the point where the angular resolution of the instrument is no longer sufficient to separate them. An often used definition of classical confusion is that it occurs when the number of synthesised beam solid angles per source falls below some value (typically 10--20). There is however no formal way to determine the level at which it affects an image. The classical confusion limit depends not only on the angular resolution of the image but its depth (coupled to the shape of the source counts function), as well as local astronomical clustering and sample variance effects \citep{heywood2013}.
For example, the RMS of the pixels in the central region of the (robust 0.0) residual COSMOS image is $\sim$5.5 $\mu$Jy~beam$^{-1}$, although this will contain some contribution from the sub-5.5$\sigma$ sources that did not get included in the final cleaning mask, were not deconvolved, and thus are still present in the residual image. This is therefore likely an overestimate of the true confusion limit.

\begin{figure*}
\centering
\includegraphics[width=6.8in]{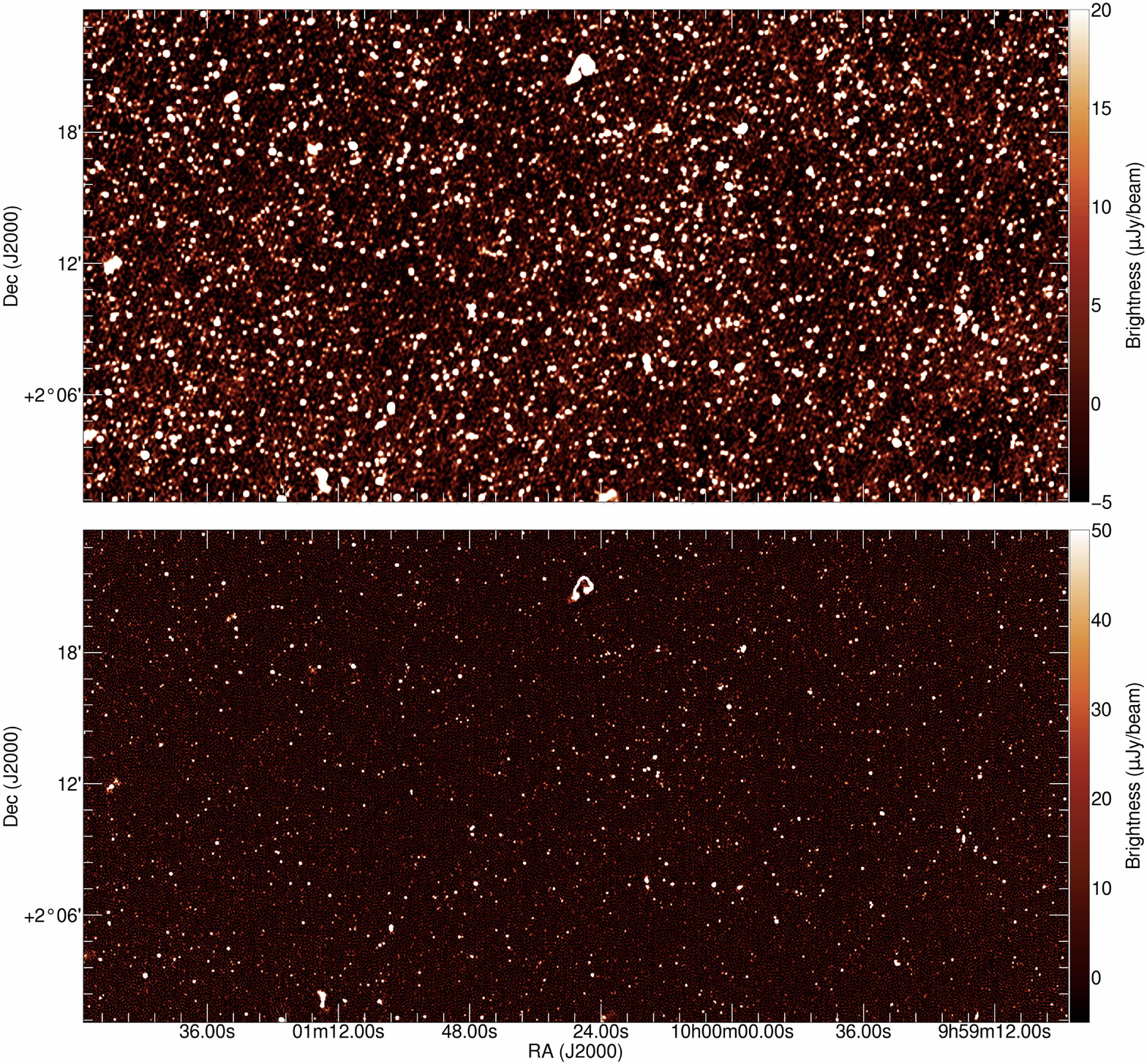}
\caption{The central 0.75~$\times$~0.375 deg$^{2}$ of the COSMOS Early Science image, showing the effects of the two different weighting schemes. The upper and lower images have angular resolutions of 8$''$.6 and 5$''$ respectively. The higher angular resolution image is noise limited rather than confusion limited, with a 1$\sigma$ noise level of 5.5 $\mu$Jy beam$^{-1}$.}
\label{fig:zooms}
\end{figure*}

Figure \ref{fig:zooms} shows the inner 0.75 $\times$ 0.375 deg$^{2}$ of the COSMOS image, in order to contrast the image in Figure \ref{fig:cosmos} with its higher angular resolution, less confused counterpart. The higher angular resolution (robust $-$1.2) images in COSMOS and XMM-LSS are not limited by classical confusion, and have 1$\sigma$ noise levels of 5.5 and 6 $\mu$Jy beam$^{-1}$ respectively. The angular resolution of the robust $-$1.2 images is 5$''$ in both of the Early Science fields.

\subsection{Effective frequency images}
\label{sec:effectivefreq}

The antenna primary beam response causes the gain of a telescope to be a strong function of direction, but the pattern width is also inversely proportional to the observing frequency. Although MFS imaging techniques allow broadband interferometric data to be used to produce deep continuum images, the effective frequency at which each pixel samples the sky brightness distribution in the final image can differ significantly from the nominal band centre frequency. A source catalogue derived from a broadband image or mosaic will contain flux density and brightness measurements that are made at a range of effective observing frequencies. Here we determine the effective observing frequency as a function of position in order to provide consistent measurements in the source catalogues (see Section \ref{sec:catalogues}).

For each individual pointing, we calculate the effective frequency (for a spectrally flat source) $\nu_{\mathrm{eff}}$ for each pixel ($x$,$y$) in the full-band image using the weighted mean
\begin{equation}
\nu_{\rm eff} (x,y) = \frac{\sum_{i} \nu_i~A_{i}(x,y)~\sigma_{i}^{-2}}{\sum_i A_{i}(x,y)~\sigma_{i}^{-2}},
\end{equation}
where $A_{i}(x,y)$ is the primary beam attenuation at pixel $(x,y)$ for sub-band $i$, and $\sigma_{i}$ is the 1$\sigma$ noise level in sub-band $i$. The noise levels are measured directly from each of the eight sub-band images. It is assumed that these sub-band sensitivity measurements coarsely capture several effects, including the differing amount of RFI losses in each sub-band, the frequency-dependent system temperature, and the imaging weights. The effective frequency will also be sensitive to the latter, since the robust $-$1.2 images will give more weight to longer baselines, and RFI generally affects shorter spacings. The relative contribution of RFI-prone regions will thus be decreased in images made with weighting that is closer to uniform.

For the XMM-LSS field, the three per-pointing effective frequency maps are mosaicked together using the same weighting scheme as was used for the total intensity mosaic shown in Figure \ref{fig:xmmlss}. The suitability of using a narrowband primary beam correction factor in a broadband MFS image is a separate issue that we investigate in Section \ref{sec:primarybeams}.

\begin{figure}
 \includegraphics[width=\columnwidth]{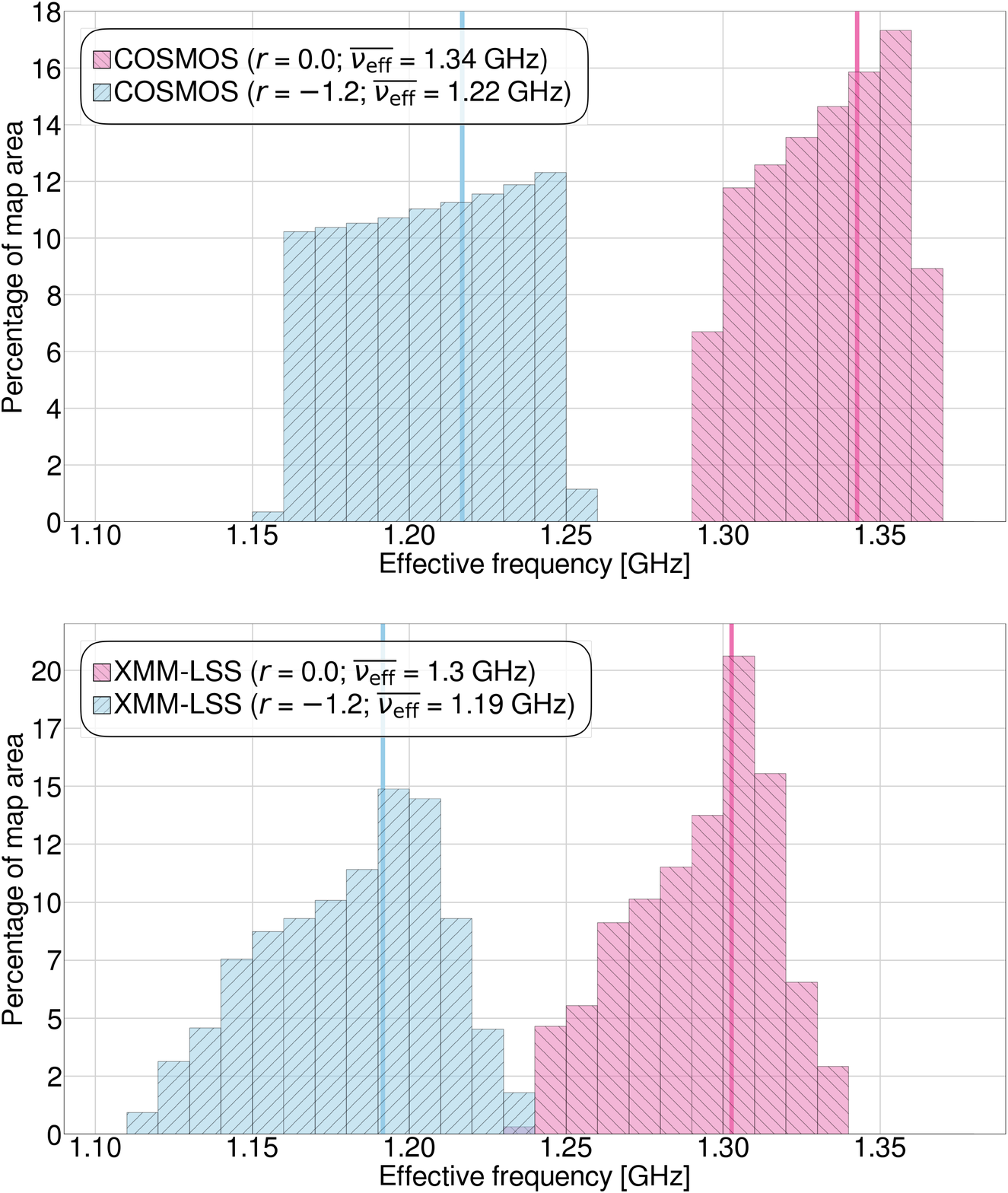}
 \caption{Histogram of the effective frequency values as a function of area for the COSMOS pointing (upper panel) and the XMM-LSS mosaic (lower panel). The two plots per panel show the distributions for the two different robust ($r$) values used in the imaging. The range of effective frequencies present in both images spans approximately 100~MHz. The vertical lines show the mean effective frequencies of each of the four images.}
 \label{fig:nu_eff}
\end{figure}

Figure \ref{fig:nu_eff} shows histograms of the effective frequency distribution as a function of area for both the COSMOS and XMM-LSS images, and for each of the two robust values that the data were imaged with. The weighting that delivers higher angular resolution has a lower effective frequency, consistent with the distribution of the RFI across MeerKAT's L-band, and it affecting primarily shorter spacings. The range of effective frequencies in the mosaic spans approximately 100~MHz in all cases, and deviates from the nominal band centre frequency by a maximum of 157~MHz (the lowest effective frequency in the high resolution XMM-LSS mosaic). This corresponds to a maximum flux density correction factor of 8.4 per cent for a source with a spectral index of $-$0.7. We return to this issue in the context of the Early Science source catalogues in Section \ref{sec:frequency_corrections}.

\subsection{Radio component catalogues}
\label{sec:catalogues}

Here we describe the production of the catalogues for the MIGHTEE Early Science continuum data. This process makes use of an automated source finder to extract features from the image. The resulting database of components is then processed further, making use of some of the additional image products described above. We refer to the `raw' output of the source finding software as the Level-0 catalogue, and the processed catalogue as the Level-1 catalogue, both of which are made available. The structure of the Level-1 catalogue is given in Appendix \ref{sec:cataloguestructure}. The cross-identification of the radio components with their optical / NIR counterparts (Level-2 catalogues) will be presented by Prescott et al.~(\emph{in prep.}).

\subsubsection{Source finding}
\label{sec:source_finding}

We used the {\sc pybdsf} source finder \citep{mohan2015} to locate and characterise components in the primary beam corrected COSMOS image, and the XMM-LSS mosaic. Briefly, {\sc pybdsf} works by using a sliding box to estimate the local background noise ($\sigma_{\mathrm{local}}$) as a function of position in the image. It then locates pixels in the image whose brightness exceeds the local background noise by some factor (in this case 5$\sigma_{\mathrm{local}}$). A flood-fill algorithm is then used to identify islands of contiguous emission down to some secondary threshold, in this case 3$\sigma_{\mathrm{local}}$. These islands are then iteratively fitted with point and 2D Gaussian components, and {\sc pybdsf} then attempts to group point and Gaussian components into sources, based on the brightness of the pixels between components in relation to the secondary threshold, and a separation criterion based on the measured sizes of the components. A catalogue describing the properties of the components and source groupings is then exported, and this raw output forms our Level-0 catalogue products. The Level-0 catalogues derived from the robust 0.0 images contain 9,915 and 20,397 components in COSMOS and XMM-LSS respectively. The steps in the sections that follow describe the modifications and additions that are made to produce the Level-1 catalogues.

\subsubsection{Visual inspection}

We visually examined the components by overlaying them on the FITS images for COSMOS and XMM-LSS. Spurious features around strong sources were removed. The nearby (z~=~0.007651, see also \citealt{maddox2021}) spiral galaxy NGC 895 (RA 02\hhh21\mmm36\fs47 -05\ddd31\dmm17\farcs0) in XMM-LSS is also bisected by the edge of the mosaic where the primary beam cut-off was applied, so the incomplete set of components associated with this source was also removed from the catalogue. In total, 19 and 123 components were removed from COSMOS and XMM-LSS respectively, bringing their total numbers of components to 9,896 and 20,274. Note that although {\sc pybdsf} makes some attempt to group \emph{components} together into \emph{sources}, the results are generally not perfect. Users of the catalogue products who wish to detemine the properties of extended sources that consist of many components should perform this grouping manually. Prescott et al.~(\emph{in prep.}) will provide a Level-2 catalogue that groups the radio components of a particular source together, in addition to identifying the optical hosts of the MIGHTEE continuum detections.

\subsubsection{Resolved sources}
\label{sec:resolved}

Following \citet{murphy2017}, the deconvolved size ($\theta_{M}$~$\times$~$\theta_{m}$) of a Gaussian component with a fitted size of ($\phi_{M}$~$\times$~$\phi_{m}$), where the $M$ and $m$ subscripts denote the major and minor axes respectively, is given by\begin{equation}
\label{eq:deconvolved}
\theta_{M} = \sqrt{\phi_{M}^{2} - \theta_{\mathrm{beam}}^{2}}
\end{equation}
where $\theta_{\mathrm{beam}}$ is the FWHM of the circular restoring beam (and similarly for $\theta_{m}$ and $\phi_{m}$; see \citealt{wild1970} for the equations for the generalised case involving an elliptical restoring beam). Dropping the $M$ and $m$ subscripts, the uncertainty in the deconvolved source size is calculated according to
\begin{equation}
\label{eq:uncertainties}
\sigma_{\theta} = \sigma_{\phi} \left[ 1 - \left(\frac{\theta_{\mathrm{beam}}}{\phi}\right)^{2}\right]^{-1/2}.
\end{equation}
where $\sigma_{\phi}$ is the uncertainty in the fitted image component, as returned by {\sc pybdsf}. A source is deemed to be reliably resolved if its deconvolved major axis size exceeds the FWHM of the restoring beam by
\begin{equation}
\label{eq:resolved}
\phi_{M} - \theta_{\mathrm{beam}} \geq 2\sigma_{\phi_{M}}.
\end{equation}
For each component in the Level-0 catalogue we compute the deconvolved source sizes and associated uncertainties according to Equations \ref{eq:deconvolved} and \ref{eq:uncertainties}, and evaluate the relationship given in Equation \ref{eq:resolved} to flag sources that are deemed to be reliably resolved. The resulting values are provided in the Level-1 catalogues. For sources where {\sc pybdsf} returns unphysical fitted source sizes (i.e. the major or minor axes of the fitted components have extents that are smaller than that of the restoring beam), the Level-1 catalogues contain zeros for the deconvolved component sizes. The number of components that are flagged as resolved are 898 (10\%) and 1376 (7\%) in COSMOS and XMM-LSS respectively.

\subsubsection{Effective frequencies}
\label{sec:frequency_corrections}

For each component in the catalogue we extracted the effective frequency ($\nu_{\mathrm{eff}}$) at its position from the maps described in Section \ref{sec:effectivefreq}, assuming a spectral index of $-$0.7, in order to correct the flux density and peak brightness measurements to a common frequency of 1.4~GHz. This frequency is selected as it is the commonly used reference frequency for previous L-band continuum studies. As noted in Section \ref{sec:effectivefreq} this correction factor will be at most 8.4\% for the canonical synchrotron spectral index of $-$0.7. The mean correction factor across both Early Science fields is 4\%, with a standard deviation of 1\%.

\section{Discussion}
\label{sec:discussion}

Parts of the following subsections rely on the cross-matching of the component catalogues with additional catalogues, either from observations made with the VLA, or with those derived from alternative processing (see Section \ref{sec:primarybeams}) of the MIGHTEE data. This is primarily for validation of the MIGHTEE continuum data products. In all cases where a cross-match is performed we conduct a nearest-neighbour match, and enforce the criteria that the radial separation of the two components must be less than 1$''$, and the components in both catalogues must be described only by a single point or Gaussian component. The latter is to minimise any biases introduced by differing angular resolution, or sensitivity to extended emission. In the case where we make use of VLA data, rather than using the published catalogues, we obtain the radio images and run {\sc pybdsf} on them as described in Section \ref{sec:source_finding} in order to further ensure that the catalogues being matched are as consistent as possible.

\subsection{Astrometry verification}
\label{sec:astrometry}

\begin{figure}
 \includegraphics[width=\columnwidth]{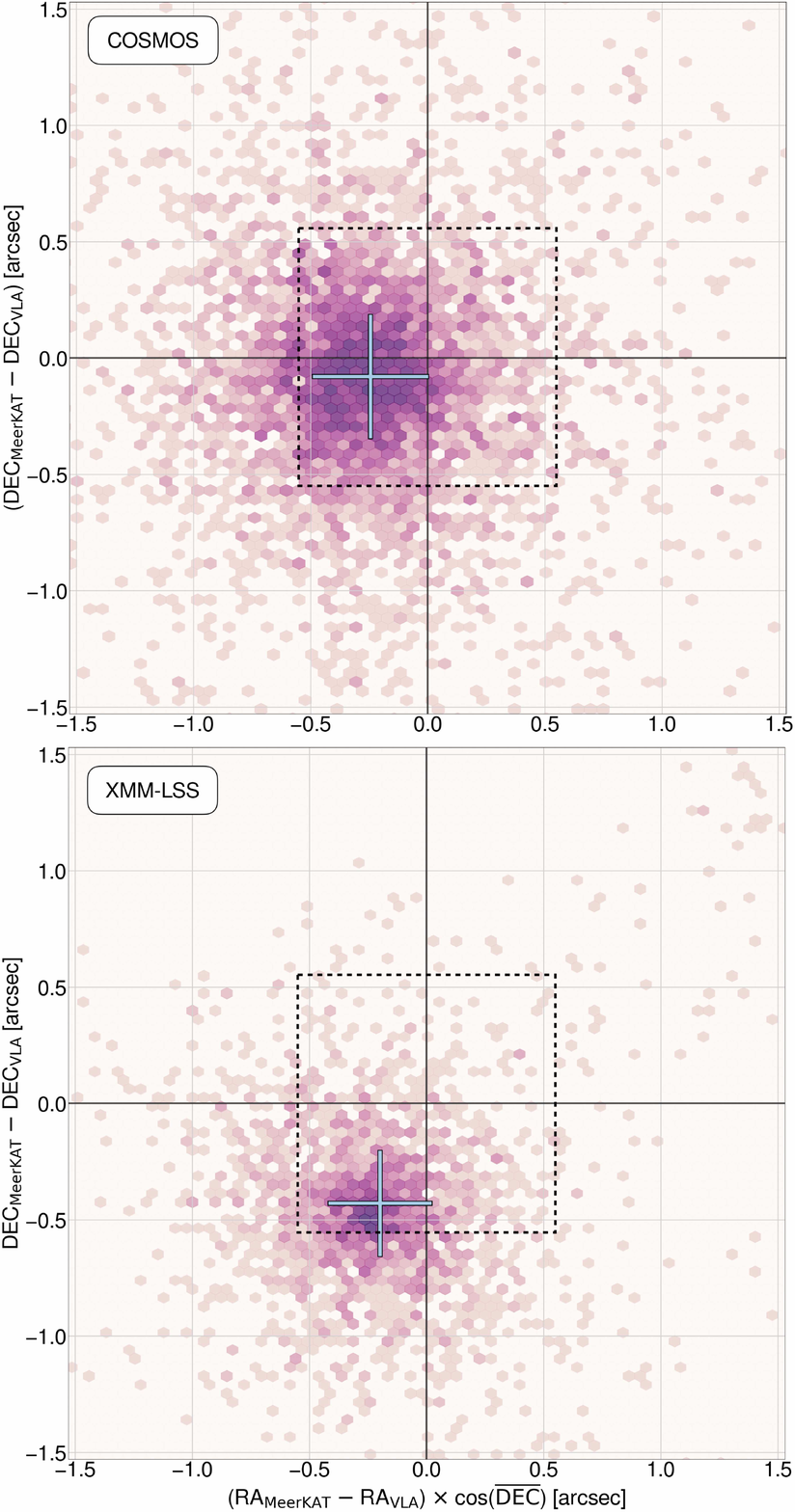}
 \caption{Differences between the MeerKAT and VLA positions for the matched components as described in Section \ref{sec:astrometry}. The figures show a 2D log-scale histogram of the distribution for COSMOS and XMM-LSS, and the cross shows the mean offset values, with the extent of the cross being 1$\sigma$ in both directions. The mean (RA,Dec) values are (-0$''$.27 $\pm$ 0$''$.01, -0$''$.19 $\pm$ 0$''$.01) and (-0$''$.20 $\pm$ 0$''$.01, -0$''$.43 $\pm$ 0$''$.01) for COSMOS and XMM-LSS respectively, with the uncertainties quoted here being the standard error of the mean. The dashed box in the centre of the Figure shows the extent of a single map pixel (1\farcs1 $\times$ 1\farcs1) in the MIGHTEE data.}
 \label{fig:astrometry}
\end{figure}

The absolute positional accuracy of a radio source is determined by a statistical component that is related to the signal-to-noise ratio (SNR) of the source and the angular resolution \citep[e.g.][]{condon1997}, as well as a systematic component that is governed principally by the accuracy of the calibrator positions, and the transfer of the gain solutions derived from them. Atmospheric effects and primary beam phase gradients can also play a role in the wide field regime.

Determining the positional accuracy of the MIGHTEE data is important, as the potential of the data is fully realised when it is combined with observations of these fields from other facilities, including radio observations at other frequencies. Early observations made with MeerKAT had some potential issues that caused systematic errors in the measured positions of sources \citep[][see also Knowles et al., \emph{submitted}]{mauch2020}. The primary effects were: (i) timestamp offsets of 2 s (one correlator-beamformer interval), causing $u,v,w$ coordinate errors that manifested as an apparent rotation about the phase centre; (ii) erroneous pointing (or catalogued) positions for calibrator sources, introducing positional offsets corresponding to that error into the target field if the calibrator was assumed to be a at the phase centre. A third consideration is the presence of significant structure in the emission of either the calibrator itself, or in the surrounding field. The effect of this has on the gain solutions is qualitatively similar to case (ii) above.

The first issue above was fixed by recomputing the $u,v,w$ coordinates in the first stages of the data processing, and the MIGHTEE data were not ostensibly affected by the second issue. We investigate any remaining unforeseen astrometric issues by comparing the positions of the MIGHTEE sources to those measured from two surveys using the VLA, both of which have superior angular resolution to MeerKAT. In COSMOS we make use of the VLA-COSMOS 3~GHz Large Project \citep{smolcic2017} with an angular resolution of 0$''$.75, and in XMM-LSS the positional measurements of a 1.5~GHz survey with 4$''$.5 angular resolution \citep{heywood2020a}. Both of these VLA surveys are shallower than the robust 0.0 MIGHTEE Early Science images, however they cover slightly larger areas in both fields. 

Figure \ref{fig:astrometry} shows the differences in RA and Dec between the MeerKAT and VLA components for COSMOS (upper panel) and XMM-LSS (lower panel). The mean offsets in (RA, Dec) are  (-0$''$.27 $\pm$ 0$''$.01, -0$''$.19 $\pm$ 0$''$.01) for COSMOS, and (-0$''$.20 $\pm$ 0$''$.01, -0$''$.43 $\pm$ 0$''$.01) for XMM-LSS, with the uncertainties quoted being the standard error of the mean. The positions of the mean offsets are marked on the figure by the blue `+' symbol, the extent of which shows the standard deviation of the underlying offset distribution. 

The forthcoming Level-2 catalogues (Prescott et al., \emph{in prep}) contain host galaxy information from the visual cross-matching of the MIGHTEE Early Science data with the optical/NIR data \citep[the latter being tied to the GAIA DR2 astrometric frame;][]{gordon2021}. This process reveals mean offsets between the MIGHTEE positions and their optical/NIR hosts of (-0$''$.03, 0$''$.01) in (RA, Dec) in the COSMOS field. But users of the MIGHTEE Early Science data should be aware that sub-pixel astrometric errors may be present between the MIGHTEE data and an external data set.

\subsection{Photometry verification}
\label{sec:photometry}

Here we validate the integrated flux density measurements from the MIGHTEE Early Science catalogues (Section \ref{sec:catalogues}) by comparing them to matched components measured with the VLA. For the COSMOS field we use the
1.4~GHz measurements from the VLA-COSMOS Survey \citep{schinnerer2010} which was observed with a combination of the extended A and compact C configurations of the VLA. While this has significantly higher angular resolution (1$''$.5) than MIGHTEE, we opt for this over the deeper VLA-COSMOS 3~GHz Large Project \citep{smolcic2017} as the observing frequency is much closer to that of MIGHTEE and reduces the uncertainty introduced by spectral index corrections. Similarly, to compare the integrated flux density measurements in the XMM-LSS field we use a VLA mosaic at L-band made from observations in the more compact C and D configurations \citep{peters2019}. This mosaic uses the same pointing grid as the B-configuration mosaic used in Section \ref{sec:astrometry} \citep{heywood2020a}. Although the depth is shallower than the B-configuration mosaic, the angular resolution is lower, and the short spacing coverage is much better than that of the B-configuration mosaic.

\begin{figure}
 \includegraphics[width=\columnwidth]{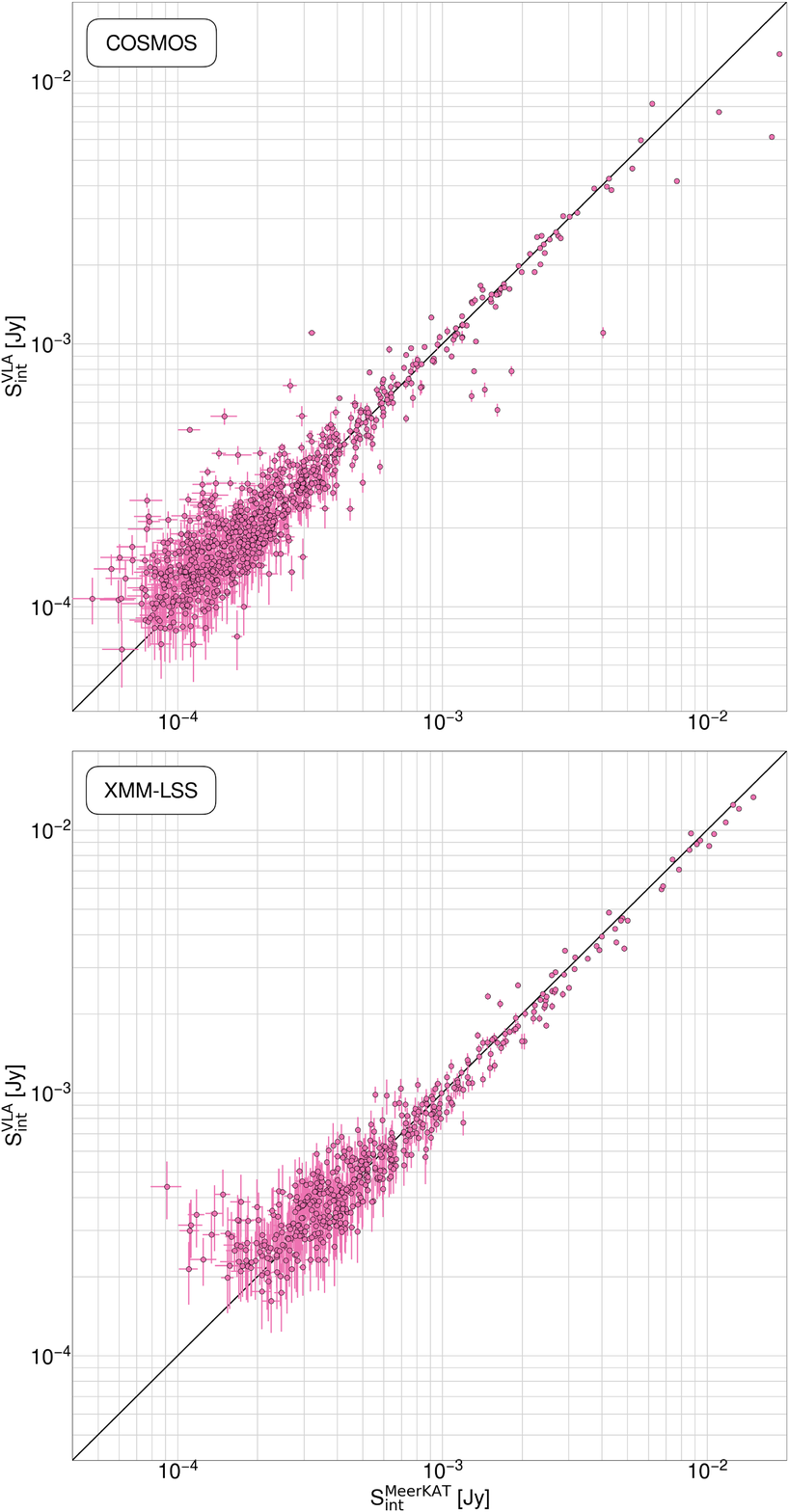}
 \caption{Integrated flux density measurements from the MIGHTEE catalogues plotted against matched components from VLA observations of the corresponding field. Details of the matching procedure is given in Section \ref{sec:discussion}, and of the VLA observations used in Section \ref{sec:photometry}. The 1:1 line is shown across the diagonal of the figure.}
 \label{fig:cosmos_photometry}
\end{figure}

The cross-matched VLA and MeerKAT components are shown on Figure \ref{fig:cosmos_photometry} for the COSMOS and XMM-LSS fields. The increasing scatter with decreasing integrated flux density due to the increasing fractional noise contribution is evident. The diagonal line on the plot shows the 1:1 ratio. The (mean,median) MeerKAT:VLA ratios for the entire distribution are (0.96,0.95) for COSMOS and (0.99,1.0) for XMM-LSS, and are therefore consistent.

\subsection{Reliability of resolved sources}
\label{sec:resolved_reliability}

We estimate the reliability of the resolved criterion described in Section \ref{sec:resolved} by means of a simulation. A true point source is injected at a random location into the COSMOS image, with peak brightness drawn from a uniform distribution with values between 10 $\mu$Jy beam$^{-1}$ and 1 mJy beam$^{-1}$. The {\sc pybdsf} source finder is run on the resulting image, and the properties of the fitted component at the location of the injected source are compared. We repeated this process to compare the true vs recovered properties of 82,000 simulated sources.

\begin{figure}
 \includegraphics[width=\columnwidth]{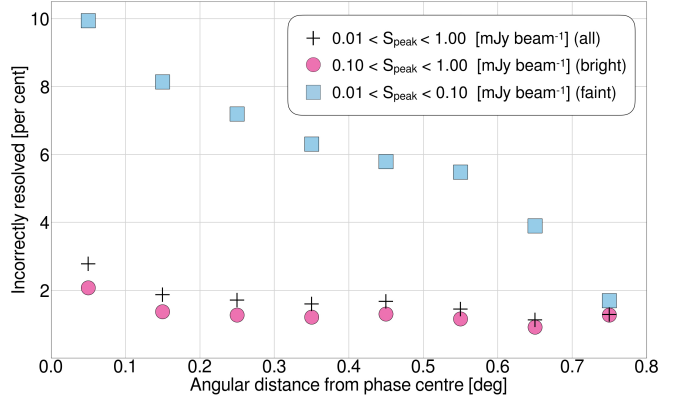}
 \caption{Percentage of sources that are erroneously marked as resolved as a function of distance from the phase centre in the COSMOS image. The results are presented in two peak brightness bins, as well as the total counts, as indicated on the legend. This plot was generated by injecting a true point source into the data and recovering its properties using {\sc pybdsf}, a process that we repeated to characterise 82,000 simulated sources.}
 \label{fig:resolved}
\end{figure}

Figure \ref{fig:resolved} shows the percentage of true point sources that are (erroneously) deemed to be resolved as a function of angular distance from the phase centre of the COSMOS image, and in two peak brightness bins (0.01 $<$ $S_{\mathrm{peak}}$ $<$ 0.1, and 0.1 $<$ $S_{\mathrm{peak}}$ $<$ 1.0, with peak brightness values in mJy beam$^{-1}$). In the higher brightness bin the fraction of sources that are erroneously found to be resolved is 1--2 per cent. For the lower brightness bin the fraction reaches 10 per cent in the centre of the map, with a gradual decline to 2 per cent as the angular distance increases. The thermal noise in the MIGHTEE Early Science images is below 2 $\mu$Jy beam$^{-1}$, and the source finder was configured with a peak brightness threshold of 5$\sigma_{\mathrm{local}}$. Thus although the faintest sources in the simulation have peak brightnesses that are $\sim$5 times the thermal noise, in practice these sources are never recovered: classical confusion elevates the effective noise in the centre of the image, and the primary beam correction raises both the confusion and thermal contributions away from the field centre. The shape of the curve for the lowest brightness bin will be governed by these two effects. Although the simulated source brightness was drawn from a uniform distribution between the two limits, the brighter sources dominate the counts in the simulation. This explains why the fractions for the total simulated population shown on Figure \ref{fig:resolved} is closer to those of the higher brightness bin. We investigate the issue of completeness in the next section.

Since the goal for the final MIGHTEE continuum survey is to achieve uniform sensitivity by means of close-packed mosaics, the innermost angular separation bin of Figure \ref{fig:resolved} is likely to be representative for the fraction of reliably resolved sources. In summary, in regions where the survey reaches its target depth, the fraction is conservatively 3 per cent on average, but may rise to 10 per cent for the fainter source population.

\subsection{Completeness}
\label{sec:completeness}

\begin{figure}
 \includegraphics[width=\columnwidth]{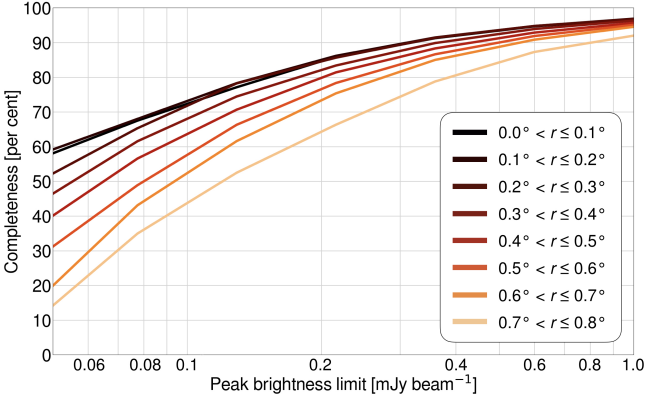}
 \caption{Point source completeness (expressed as a percentage) as a function of peak brightness in the COSMOS image. The completeness percentages are determined in concentric annuli with angular separation ranges from the phase centre indicated on the figure legend. The uppermost curves are representative of the point source completeness for the maximum depth (confused) regions of the COSMOS image (or mosaicked image products).}
 \label{fig:completeness}
\end{figure}

We make use of the simulation described in Section \ref{sec:resolved_reliability} to examine catalogue completeness, as shown in Figure \ref{fig:completeness}. To generate this figure we simply count sources that are injected into the data but not recovered by {\sc pybdsf}. In the full-depth (confused) regions, MIGHTEE's continuum data is 96 per cent complete at 1 mJy beam$^{-1}$, 90 per cent complete at 300 $\mu$Jy beam$^{-1}$, and 60 per cent complete at 50 $\mu$Jy beam$^{-1}$.

A companion paper (Hale et al., \emph{in prep.}) presents an analysis of the differential source counts and sky background temperature, and includes a thorough analysis of the bias corrections required to account for completeness and other effects in the source counts derived from the Early Science catalogues.

\subsection{Primary beam correction in broadband, widefield imaging}
\label{sec:primarybeams}

The `traditional' approach for primary beam correction is to divide the final continuum image by an image of the primary beam pattern. For modern broadband observations the primary beam image is usually evaluated at the nominal central frequency of the observing band. This approach is potentially problematic for several reasons. The main issue, already touched upon in Section \ref{sec:effectivefreq}, is the large fractional bandwidth. In this case a single-frequency primary beam model becomes inaccurate for off-axis sources in a way that depends on both the bandwidth (including gaps due to RFI) and the intrinsic spectra of the sources. The second issue is that due to the telescope mount the primary beam is time dependent as well as frequency and direction dependent, and effects such as pointing errors impart an effective primary beam that differs from antenna to antenna. For anything other than a snapshot observation, an image-based primary beam model will also be a time (and thus direction) averaged product.

The two main alternatives to the traditional approach for primary beam correction are A-projection, and facet-based corrections. In the former approach a primary beam correction is applied to each visibility in the form of a convolution kernel during gridding \citep{bhatnagar2013}. The advantage of this approach is that in principle the frequency, time, polarisation, and direction dependencies of the primary beam can be fully modelled, however the computational cost is significant. At the time of writing, an A-projection implementation for general use with MeerKAT is forthcoming \citep{sekhar2021}, so this is not tested here. Facet-based primary beam corrections can be applied by {\sc ddfacet} \citep{tasse2018} in much the same way as the direction-dependent gain corrections that are applied on a per-tessellation basis as described in Section \ref{sec:processing}. The beam model in this case consists of a multi-dimensional FITS model that contains a full 2$\times$2 Jones matrix model of the MeerKAT beam as a function of direction, evaluated at several frequency intervals. The sky image is divided up into a regular grid of facets, and a piecewise primary beam correction is evaluated for the direction corresponding to the facet centre. The beam model for each facet is averaged in time according to the parallactic angle interval of the observation, and frequency dependence is captured by evaluating and applying the beam model for each of the (de-)gridding sub-bands. The beam corrections can also be optionally smoothed in the spatial dimensions. The advantage of this approach is that the sky model during deconvolution is generated in an intrinsic rather than apparent form, allowing for constraints on the source spectra that are astrophysically rather than instrumentally motivated, as well as fewer degrees of freedom when subsequently solving for direction-dependent effects.

In this section we compare the traditional and facet-based approaches, making use of additional MIGHTEE pointings in the COSMOS field, in order to match attenuated and corrected sources with counterparts that are on-axis in the additional pointings, and thus represent the intrinsic properties. 

\subsubsection{Production of additional images and catalogues}
\label{sec:additional_proc}

The MIGHTEE COSMOS observations form a close-packed mosaic of 15 individually deep pointings, the arrangement of which can be seen in Figure \ref{fig:schematic}. The inner regions of the numbered pointings where the primary beam response is relatively flat can be used to provide intrinsic integrated flux density and peak brightness measurements for ensembles of sources that lie off-axis to varying degrees in the single central Early Science pointing. Thus the apparent and beam-corrected intrinsic measurements from the single pointing can be compared to a set of true intrinsic measurements derived from the additional pointings to test the validity of the primary beam model. For the true intrinsic measurements we restrict source extraction to the inner 0.12 degree radius of the 14 additional pointings. In this region the nominal primary beam gain does not fall below 97\%. 

Each of the 14 additional pointings were processed using the methods described in Section \ref{sec:obs_and_proc}, up to and including the correction for DDEs (although no peeling step was used for any COSMOS pointings). The model and residuals for each pointing were also convolved to a common resolution prior to mosaicking. No primary beam correction was applied to the individual pointings, however each of them was masked to the circular regions shown on Figure \ref{fig:schematic} prior to mosaicking with {\sc montage} in order to produce a single FITS image from which to extract a catalogue. The {\sc pybdsf} source finder was run on the resulting image. Components that were within 18$''$ of the edge of a pointing were excluded in order to avoid comparisons between sources that were partially missing due to the radial cuts. The resulting catalogue (hereafter the `intrinsic' catalogue) contains 4,828 point and Gaussian components.

The 2GC-calibrated (Section \ref{sec:di_selfcal}) Early Science COSMOS data was imaged again with {\sc ddfacet}, however for this run a model of the primary beam was used. The model was applied to the data over 1024 facets covering the 3.13 $\times$ 3.13 deg$^{2}$ area shown in Figure \ref{fig:schematic}, although the beam model is interpolated and smoothed by {\sc ddfacet} when producing the final image, with an independent beam correction evaluated in each of 10 sub-bands. The beam model was generated using the {\sc eidos} package \citep{asad2021}. The real and imaginary part of each correlation product (XX, XY, YX and YY) are captured via a FITS cube (for a total of eight cubes). Each cube covers 6 $\times$ 6 deg$^{2}$ by means of 257 $\times$ 257 spatial pixels, and the full MeerKAT L-band is captured in 95 frequency channels. 

Deconvolution of the data with the beam model enabled used the sub-space deconvolution mode (SSD2), with a second-order log-space polynomial capturing the (intrinsic) sky frequency dependence. Five major cycles were performed, with 60,000 iterations, and making use of the same post-2GC cleaning mask that was employed to produce the images described in Section \ref{sec:dd_selfcal}. 

\begin{figure}
 \includegraphics[width=0.95 \columnwidth]{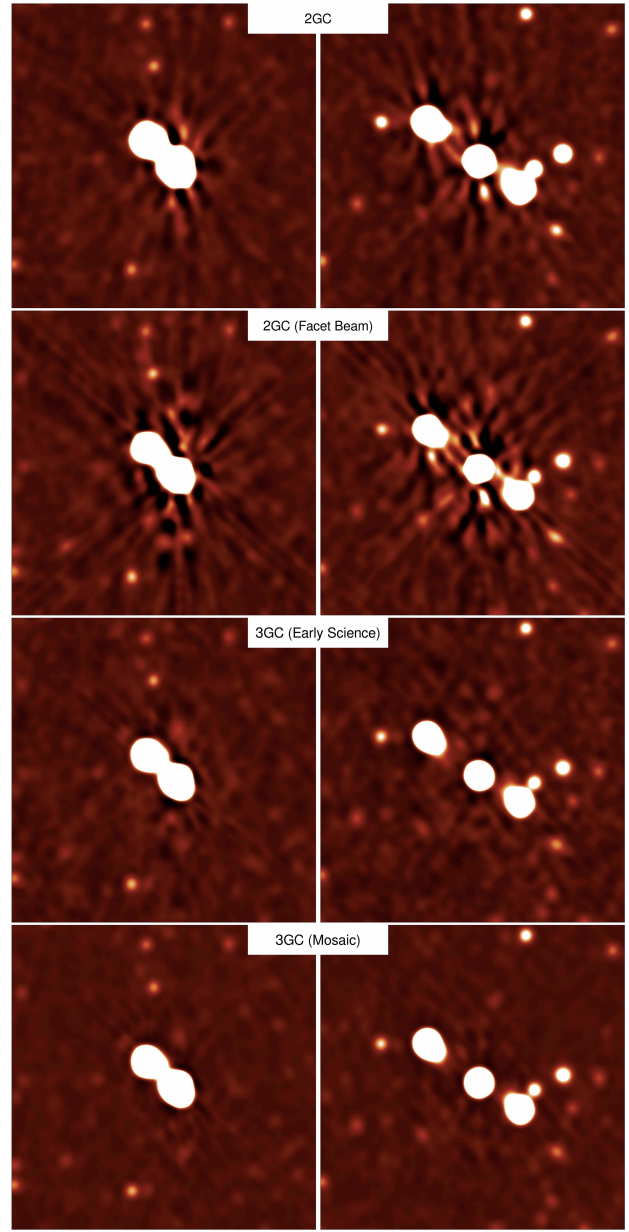}
 \caption{Thumbnails spanning 4$'$, showing two bright double radio galaxies in the COSMOS field. Each of these sources is approximately half a degree from the pointing centre, and they are on opposite sides of the image. From top to bottom the panels show the imaging quality (i) after the 2GC step (Section \ref{sec:di_selfcal}); (ii) after the 2GC step but using SSD clean and a primary beam model (Section \ref{sec:additional_proc}); (iii) after the 3GC step (Section \ref{sec:dd_selfcal}); (iv) the full COSMOS mosaic, for which the increase in depth is apparent due the contributions of multiple overlapping pointings. The colour scale is the same as that used in Figures \ref{fig:cosmos}, \ref{fig:xmmlss}, and \ref{fig:zooms}, and runs from -50 to 150 $\mu$Jy beam$^{-1}$.}
 \label{fig:calsteps}
\end{figure}

The final primary beam corrected image was convolved to the common resolution using the same method as described in Section \ref{sec:pbcor}, and masked to the same circular region as the `traditional' Early Science COSMOS image. Once again the {\sc pybdsf} source finder was used, and the resulting `facet' catalogue contains 10,675 point and Gaussian components. The excess sources are mainly artefacts around strong sources that remain present despite the use of the beam model. Figure \ref{fig:calsteps} illustrates this, showing cutouts of two bright double radio sources from the COSMOS field to show the differing levels of artefacts for different processing stages / methods. Each of these sources is about half a degree from the pointing centre. Another cause of the increased artefacts in the image with the faceted beam model could be the second-order polynomial used to fit the spectra of the clean components. While likely to be `astrophysically' suitable given typical source spectra, a higher order fit may allow more flexibility to compensate for any deficiencies in the beam model for sources this far away from the pointing centre.

To produce the plots that follow, the cross-matching between the traditional (apparent and primary beam corrected), mosaic, and facet catalogues was performed using the same criteria described in Section \ref{sec:discussion}. We further impose the additional constraint that the SNR of the components must exceed 20.

\subsubsection{Apparent over intrinsic}
\label{sec:app_over_int}

\begin{figure}
 \includegraphics[width=\columnwidth]{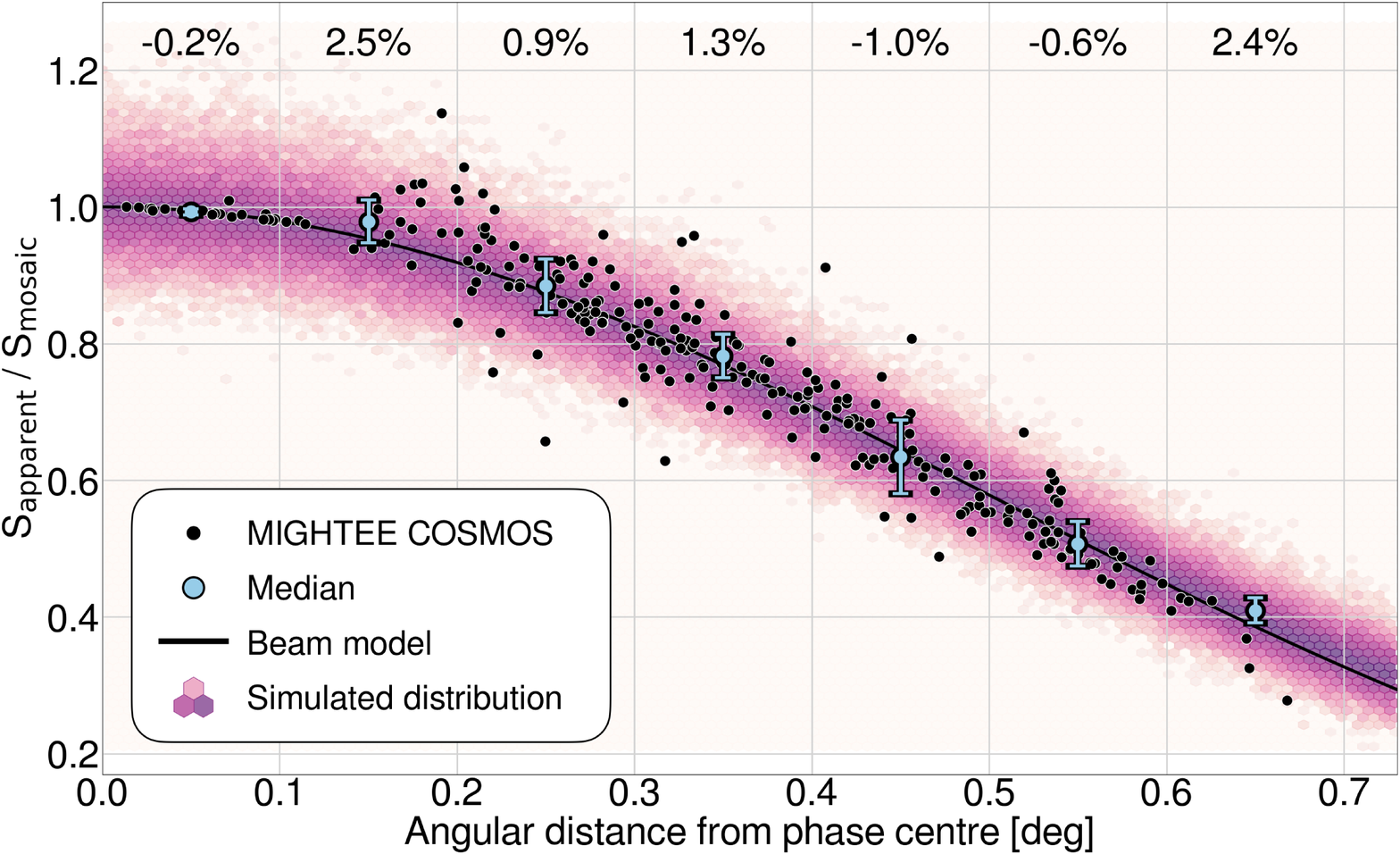}
 \caption{The apparent source flux densities from the Early Science COSMOS image divided by the presumed-intrinsic measurements produced for the `mosaic' catalogue (see Section \ref{sec:additional_proc}), as a function of the angular distance of the component from the phase centre (black points). The nominal primary beam model is shown by the solid black line, and the median (and median absolute deviation) is shown in seven bins by the heavy points with error bars. A Monte Carlo simulation is used to show the scatter on the points that is introduced by the source spectral indices and the differing noise properties of the two measurements. The results of simulating 150,000 components is shown in the background of this figure as a 2D log-scale histogram. Please refer to Section \ref{sec:app_over_int} for details. The percentages are the difference between the median points and the primary beam model.}
 \label{fig:app_over_int}
\end{figure}

An obvious check of the suitability of the primary beam model is to compare the apparent (no primary beam correction applied) flux density measurements from the MIGHTEE COSMOS Early Science image with the assumed-intrinsic measurements from the `mosaic' catalogue. Such a comparison is shown in Figure \ref{fig:app_over_int}, where the apparent to intrinsic ratio is plotted as a function of the angular distance of the source from the phase centre. The difference between the single frequency primary beam model (the solid black line) and the median measurements of the ratio in seven bins is shown on the figure, and is $\sim$1\%.

The ratio shown by the black points exhibit significant scatter, and we investigate the expected level of this scatter by means of a Monte Carlo simulation. A simulated source is assigned a radial offset in the range [0,0.7] and its true flux density is assigned by selecting a random value from the list of `mosaic' flux densities. A random spectral index value is assigned drawn from a normal distribution with a mean of $-$0.7 and a standard deviation of 0.1, consistent with the 1.4 GHz - 4.86 GHz spectral index distribution of the extragalactic sky simulation presented by \citet{wilman2008}. Using the spectral index and the true flux density, an eight-point spectrum for the simulated source is computed, representing the eight sub-bands across the L-band in which the MeerKAT data is deconvolved. This true spectrum is then perturbed with frequency-dependent primary beam attenuations, as well as thermal noise that is appropriate for each sub-band, based on the measured values that were used in Section \ref{sec:effectivefreq}. The mean value of this corrupted spectrum is then taken to be the apparent measured flux density. The simulated ratio is produced by dividing this value by the intrinsic flux density of the source perturbed by appropriate Gaussian noise. This process is repeated 150,000 times, and the resulting distribution is plotted as the 2D log-scale histogram in the background of Figure \ref{fig:app_over_int}. Note also that the additional COSMOS pointings are significantly shallower (and therefore noisier) than the central Early Science pointing (as can be see in the on-source time column of Table \ref{tab:extra_observations}). In the final COSMOS mosaic the depth will be recovered by the co-addition of the many neighbouring pointings, however that is not done for this analysis in order to minimise the effects of the antenna primary beam on the flux density measurements. The increased depth provided by the co-addition of neighbouring pointings can be seen in the lowest row of Figure \ref{fig:calsteps}.

The distribution of the measured points is well represented by the simulation. The principal cause of the scatter is the differing noise properties of the two images (illustrated further by the innermost radii where the mosaic and the Early Science image share the same data), however the spectral index of the source also plays a part via the colour correction issues mentioned in Section \ref{sec:effectivefreq}, as evidenced by the deviation between the distribution of the simulated measurements and the primary beam model at high angular separations. The distribution of spectral indices based on the \citet{wilman2008} simulation may be somewhat simplistic, e.g.~flatter spectrum AGN cores may be under-represented at these depths \citep{whittam2017}.

Despite these issues and the use of the single-frequency primary beam image for correcting broadband data, the median ratios between the apparent and intrinsic flux densities are typically $\sim$1\% and in all radius bins are better than 3\%. 

\subsubsection{Comparison of the traditional and facet-based methods}

In this final section we make a comparison between the traditional and facet-based approaches for primary beam correction, as well as comparing both of these methods to the assumed-intrinsic `mosaic' measurements. As in Section \ref{sec:app_over_int} these comparisons are made by determining the ratio of the flux densities of matched components as a function of their separation from the phase centre in order to determine the level of any systematic biases.

\begin{figure}
 \includegraphics[width=\columnwidth]{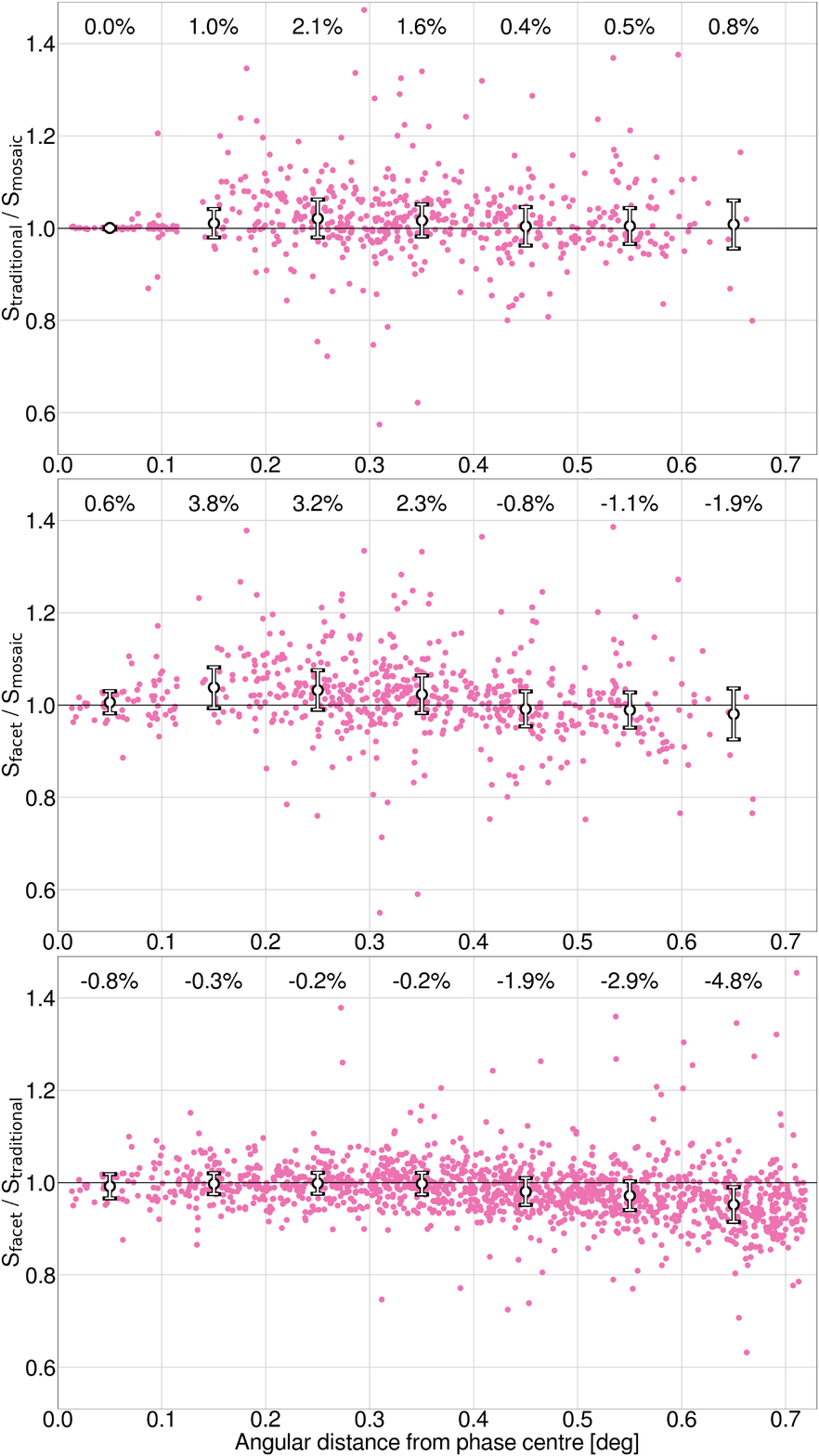}
 \caption{Flux density ratios for matched components as a function of their separation from the phase centre for (top to bottom) the traditional/mosaic, facet/mosaic, and facet/tradtional scenarios. The scatter in these plots is discussed in Section \ref{sec:app_over_int}, and the fact that the additional COSMOS pointings that form the mosaic are shallower than the Early Science pointing is evident. The median ratios (and median absolute deviations) are shown in seven bins by the points with error bars, and the deviation from unity of these points expressed as a percentage is also noted in each bin.}
 \label{fig:beam_ratios}
\end{figure}

The results are presented in Figure \ref{fig:beam_ratios}, which shows such plots for traditional/mosaic (upper panel), facet/mosaic (middle panel), and facet/traditional (lower panel). Since all the flux density measurements that go into these plots are intrinsic, the unity line represents the point where there is no difference between the methods. The level of scatter in these plots is (as expected) consistent with the scatter evident in Figure \ref{fig:app_over_int}, and as simulated in Section \ref{sec:app_over_int}. 

The median (and median absolute deviation) of the flux density ratios is plotted in seven bins, and the deviation from unity is expressed as a percentage label in each bin on the plots. The traditional correction method (upper panel) is in all bins consistent with unity. The medians of the facet/mosaic points (central panel) have hints of a sinusoidal structure, although the traditional/mosaic points also show a slight excess in the 0.1 -- 0.4 deg bins. The facet/traditional distribution also begins to deviate from unity beyond a radius of 0.4, so the real differences could mainly be in the region beyond the half power point of the primary beam. Within this region there is essentially no clear difference between the traditional and facet based methods. 

A further difference between the two methods arises via the deconvolution. As mentioned earlier, the {\sc ddfacet} SSD approach is to fit for the intrinsic properties of the sky model. When producing the final image, this model is evaluated at the reference frequency and then restored into the residual. This means that the deconvolved part of the sky will in principle not be subjected to the effective frequency issues touched upon in Sections \ref{sec:effectivefreq} and \ref{sec:photometry}. The images used in the traditional approach however, are restored with components that are the weighted mean of the MFS model, which are then restored into the MFS-weighted residual. Clearly both of these approaches have their advantages; the former avoids the primary beam and position-dependent effective frequency effects, and while the latter method is subject to these effects, the deconvolved and residual portions of the image are at least consistent. As is often the case in radio astronomy one size does not fit all, and the choice of method often depends on the science goals and the observational parameters. For survey observations it is essential to conduct (and continue to conduct) tests such as this to investigate such systematics.

\section{Conclusions}

We have described in detail the total intensity continuum processing strategy for the MeerKAT MIGHTEE survey, and validated the suitability of this approach. This has been achieved by producing the continuum Early Science data products, totalling 5.1 deg$^{2}$ in the COSMOS and XMM-LSS fields, and performing both self-consistency checks and comparisons with existing radio observations in these fields. The imaging data is presented with two different depth / angular resolution combinations, and the images are decomposed into per-field catalogues, the combined component count of which is 30,174.

Corrections for direction-dependent effects (DDEs) are essential for delivering performance-limited images from MeerKAT, a goal which is highly desirable for surveys such as MIGHTEE where uniformity of the data is key, and for maximising their legacy value when not all science goals can be foreseen. The use of a primary beam model during imaging does not fully correct for the DDEs in the image. Assuming no major deficiencies in the beam model, this suggests that the dominant causes are stochastic processes (mechanical pointing errors and possibly ionospheric effects at lower frequencies) that cannot be known \emph{a priori} and must be solved for via self-calibration techniques.

Despite the major role the antenna primary beam patterns play in these (and all other) broadband, wide-field observations, we demonstrate that the traditional method of primary beam correction (namely the division of the final science image by a single frequency beam model at the band centre) is likely to be good enough for observations of this type. It does not appear to significantly bias the photometry extracted from the images, even at significant radii from the phase centre. Flux densities extracted from the primary beam corrected COSMOS image are consistent with those extracted from the central regions of 14 additional offset pointings at the $\sim$1\% level. This level is on par with other effects that introduce photometric uncertainties, such as absolute flux-scale calibration, and the uncertain colour corrections for sources that do not have a well-constrained spectral index measurement.

Several studies have made use of these data during the development and validation phase \citep[e.g.][]{pasini2020,delhaize2021,delvecchio2021}, however a public release of the image and catalogue products that form the Early Science MIGHTEE continuum data now accompanies this article. The properties of the Early Science continuum data meet or exceed the design specifications of the MIGHTEE survey, meaning that the continuum science goals of the full survey can also be met as planned. A set of sky footprints for the MIGHTEE survey that supersede those presented along with the science overview by \citet{jarvis2016} are shown in Appendix \ref{sec:mosaics}, with machine readable pointing centres available as supplementary material.

\section*{Data availability}

The raw MeerKAT visibilities for which any proprietary period has expired can be obtained from the SARAO archive at \url{https://archive.sarao.ac.za}. The image and catalogue products presented in this article can be freely accessed from \url{https://doi.org/10.48479/emmd-kf31}. The data processing scripts used can be downloaded from \url{https://github.com/IanHeywood/oxkat} (v0.1), and the underlying software packages from the links therein.

\section*{Acknowledgements}

The MeerKAT telescope is operated by the South African Radio Astronomy Observatory, which is a facility of the National Research Foundation, an agency of the Department of Science and Innovation. We acknowledge use of the Inter-University Institute for Data Intensive Astronomy (IDIA) data intensive research cloud for data processing. IDIA is a South African university partnership involving the University of Cape Town, the University of Pretoria and the University of the Western Cape. The authors acknowledge the Centre for High Performance Computing (CHPC), South Africa, for providing computational resources to this research project. We thank Chris Schollar, Masechaba Sydil Kupa, and Fernando Camilo for SARAO archive and DOI support. IH and MJJ acknowledge support from the UK Science and Technology Facilities Council [ST/N000919/1]. IH acknowledges support from the South African Radio Astronomy Observatory which is a facility of the National Research Foundation (NRF), an agency of the Department of Science and Innovation. IH thanks the Rhodes Centre for Radio Astronomy Techniques and Technologies (RATT), South Africa, for the provision of computing facilities. CLH acknowledges support from the Leverhulme Trust through an Early Career Research Fellowship. The research of OS is supported by the South African Research Chairs Initiative of the Department of Science and Technology and National Research Foundation. JA acknowledges financial support from the Science and Technology Foundation (FCT, Portugal) through research grants PTDC/FIS-AST/29245/2017, UIDB/04434/2020 and UIDP/04434/2020. PNB is grateful for support from the UK STFC via grant ST/V000594/1. The research of RPD is supported by the South African Research Chairs Initiative (grant ID 77948) of the Department of Science and Innovation and National Research Foundation. NM acknowledges support from the Bundesministerium f{\"u}r Bildung und Forschung (BMBF) award 05A20WM4. IP acknowledges financial support from the Italian Ministry of Foreign Affairs and International Cooperation (MAECI Grant Number ZA18GR02) and the South African Department of Science and Technology's National Research Foundation (DST-NRF Grant Number 113121) as part of the ISARP RADIOSKY2020 Joint Research Scheme. This research made use of Astropy,\footnote{\url{http://www.astropy.org}} a community-developed core Python package for Astronomy \citep{astropy2013, astropy2018}. This research has made use of NASA's Astrophysics Data System. This research made use of Montage, which is funded by the National Science Foundation under Grant Number ACI-1440620, and was previously funded by the National Aeronautics and Space Administration's Earth Science Technology Office, Computation Technologies Project, under Cooperative Agreement Number NCC5-626 between NASA and the California Institute of Technology. Finally, we thank the anonymous referee and the MNRAS editorial staff for their comments, which significantly improved the final article.





\subsubsection*{Author Affiliations}
\footnotesize
$^{1}$Astrophysics, Department of Physics, University of Oxford, Keble Road, Oxford, OX1 3RH, UK\\ 
$^{2}$Centre for Radio Astronomy Techniques and Technologies, Department of Physics and Electronics, Rhodes University, PO Box 94, Makhanda, 6140, South Africa\\
$^{3}$South African Radio Astronomy Observatory, 2 Fir Street, Black River Park, Observatory, Cape Town, 7925, South Africa\\
$^{4}$Physics Department, University of the Western Cape, Private Bag X17, Bellville, 7535, South Africa\\
$^{5}$Institute for Astronomy, Royal Observatory Edinburgh, Blackford Hill, Edinburgh, EH9 3HJ\\
$^{6}$The Inter-University Institute for Data Intensive Astronomy (IDIA), Department of Astronomy, University of Cape Town, Private Bag X3, Rondebosch, 7701, South Africa\\
$^{7}$GEPI, Observatoire de Paris, CNRS, Université Paris Diderot, 5 Place Jules Janssen, 92190 Meudon, France\\
$^{8}$Instituto de Astrof\'{i}sica e Ci\^{e}ncias do Espa\c co, Universidade de Lisboa, OAL, Tapada da Ajuda, PT1349-018 Lisboa, Portugal\\
$^{9}$Departamento de F\'{i}sica, Faculdade de Ci\^{e}ncias, Universidade de Lisboa, Edif\'{i}cio C8, Campo Grande, PT1749-016 Lisbon, Portugal\\
$^{10}$SUPA, Institute for Astronomy, Royal Observatory Edinburgh, EH9 3HJ, UK\\
$^{11}$The Inter-University Institute for Data Intensive Astronomy (IDIA), Department of Astronomy, University of Cape Town, Private Bag X3, Rondebosch, 7701, South Africa\\
$^{12}$School of Science, Western Sydney University, Locked Bag 1797, Penrith, NSW 2751, Australia\\
$^{13}$CSIRO Astronomy and Space Science, PO Box 1130, Bentley, WA, 6102, Australia\\
$^{14}$Wits Centre for Astrophysics, University of the Witwatersrand, 1 Jan Smuts Avenue, Braamfontein, Johannesburg 2050, South Africa\\
$^{15}$Department of Physics, University of Pretoria, Hatfield, Pretoria 0028, South Africa\\
$^{16}$Department of Astronomy, University of Cape Town, Private Bag X3, Rondebosch 7701, South Africa\\
$^{17}$Centre for Astrophysics Research, School of Physics, Astronomy and Mathematics, University of Hertfordshire, College Lane, Hatfield, Hertfordshire AL10 9AB, UK\\
$^{18}$Faculty of Physics, Ludwig-Maximilians-Universit{\"a}t, Scheinerstr. 1, 81679 Munich, Germany\\
$^{19}$National Radio Astronomy Observatory, 520 Edgemont Road, Charlottesville, VA 22903, USA\\
$^{20}$INAF-IRA, Via P. Gobetti 101, I-40129, Italy\\
$^{21}$South African Astronomical Observatory, P.O. Box 9, Observatory 7935, Cape Town, South Africa\\
$^{22}$Southern African Large Telescope, P.O. Box 9, Observatory 7935, Cape Town, South Africa\\
$^{23}$A\&A, Department of Physics, Faculty of Sciences, University of Antananarivo, B.P. 906, Antananarivo 101, Madagascar\\
$^{24}$National Radio Astronomy Observatory, 1003 Lopezville Road, Socorro, NM 87801, USA\\
$^{25}$School of Astronomy, Institute for Research in Fundamental Sciences, P.O. Box 131, Tehran, Iran\\
$^{26}$Max-Planck-Institut f{\"u}r Astronomie, K{\"o}nigstuhl 17, D-69117 Heidelberg, Germany




\appendix 

\section{Structure of the Level-1 source catalogue}
\label{sec:cataloguestructure}

The Level-1 Early Science catalogues for COSMOS and XMM-LSS contain a subset of the columns from the raw {\sc pybdsf} output (the Level-0) catalogues, as well as some additional derivative columns. In addition they have been filtered for artefacts and incomplete sources that are cut off by the primary beam correction and mosaicking (see Section \ref{sec:catalogues} for full details). Below we describe the set of columns that form the Level-1 catalogues. The first ten rows of the COSMOS catalogue are presented in Table \ref{tab:tenrows}.\\
\\
\noindent
{\bf (0)}: An IAU-style identifier of the form JHHMMSS.SS+/-DDMMSS.S.
\\

\noindent
{\bf (1), (2)}: The J2000 Right Ascension of the component in degrees, and its associated 1$\sigma$ positional uncertainty.
\\

\noindent
{\bf (3), (4)}: The J2000 Declination of the component in degrees, and its associated 1$\sigma$ positional uncertainty. Note that columns (2) and (4) are the statistical uncertainties derived from the component fitting by {\sc pybdsf} and thus do not include astrometric errors. Note also that columns (1) and (3) have not been corrected for any systematic offsets. Please refer to Section \ref{sec:astrometry} for further details.
\\

\noindent
{\bf (5), (6)}: The integrated flux density and its associated 1$\sigma$ uncertainty in Jy.
\\

\noindent
{\bf (7), (8)}: The peak brightness and its associated 1$\sigma$ uncertainty in Jy beam$^{-1}$.
\\

\noindent
{\bf (9)}: The effective frequency in Hz at which the source is observed assuming a spectral index of $-$0.7 (see Section \ref{sec:effectivefreq} for details).
\\

\noindent
{\bf (10), (11)}: The integrated flux density and its associated 1$\sigma$ uncertainty in Jy, corrected to 1.4 GHz assuming the effective frequency given in column {\bf (9)} and a spectral index of $-$0.7.
\\

\noindent
{\bf (12), (13)}: The peak brightness and its associated 1$\sigma$ uncertainty in Jy beam$^{-1}$, corrected to 1.4 GHz assuming the effective frequency given in column {\bf (9)} and a spectral index of $-$0.7.
\\

\noindent
{\bf (14), (15)}: The major axis and associated 1$\sigma$ uncertainty of the 2D Gaussian fitted to the source by {\sc pybdsf}, in degrees.
\\

\noindent
{\bf (16), (17)}: The minor axis and associated 1$\sigma$ uncertainty of the 2D Gaussian fitted to the source by {\sc pybdsf}, in degrees.
\\

\noindent
{\bf (18), (19)}: The position angle (measured east of north) and associated 1$\sigma$ uncertainty of the 2D Gaussian fitted to the source by {\sc pybdsf}, in degrees.
\\

\noindent
{\bf (20), (21)}: The major axis and associated 1$\sigma$ uncertainty of the deconvolved source size, in degrees, evaluated according to Equations \ref{eq:deconvolved} and \ref{eq:uncertainties}, see Section \ref{sec:resolved} for details. A zero in these columns means that the source was fitted with an unphysical size by {\sc pybdsf} and is assumed to be unresolved along the major axis.
\\

\noindent
{\bf (22), (23)}: The minor axis and associated 1$\sigma$ uncertainty of the deconvolved source size, in degrees, evaluated as per columns {\bf (19)} and {\bf (20)}. A zero in these columns means that the source was fitted with an unphysical size by {\sc pybdsf} and is assumed to be unresolved along the minor axis.
\\

\noindent
{\bf (24)}: A Boolean flag indicating if the source is reliably resolved, i.e. if it satisfies the criterion given in Equation \ref{eq:resolved}.
\\

\noindent
{\bf (25)}: The estimate from {\sc pybdsf} of the background RMS around the component, in Jy beam$^{-1}$.
\\

\noindent
{\bf (26)}: A unique integer identifier for the Gaussian component from the raw {\sc pybdsf} output.
\\

\noindent
{\bf (27)}: A unique integer identifier for the source from the raw {\sc pybdsf} output.
\\

\noindent
{\bf (28)}: A unique integer identifier for the island from the raw {\sc pybdsf} output. Columns 25, 26 and 27 are provided in the Level 1 catalouges to preserve the source finder's original groupings of Gaussian components into sources.
\\

\begin{table*}
\begin{minipage}{176mm}
\centering
\caption{The first ten rows from the COSMOS Level 1 Early Science catalogue. Please refer to the text in Appendix \ref{sec:cataloguestructure} for full column descriptions.}
\label{tab:tenrows}
\begin{tabular}{llllllllll} \hline
 & Name & RA & $\sigma_{\mathrm{RA}}$ & Dec & $\sigma_{\mathrm{Dec}}$ & $S_{\mathrm{int}}$ & $\sigma_{\mathrm{local}}$ & $S_{\mathrm{peak}}$ & $\sigma_{\mathrm{peak}}$ \\
 & & (deg) & (deg) & (deg) & (deg) & (Jy) & (Jy) & (Jy beam$^{-1}$) & (Jy beam$^{-1}$) \\
 & (0) & (1) & (2) & (3) & (4) & (5) & (6) & (7) & (8) \\ \hline
\textcolor{mygray}{\emph{1}} & J100320.73+020931.7 & 150.83641 & 0.00029 & 2.15882 & 0.00030 & 0.0001038 & 0.0000289 & 0.0000555 & 0.0000107 \\
\textcolor{mygray}{\emph{2}} & J100320.26+021312.7 & 150.83442 & 0.00003 & 2.22022 & 0.00008 & 0.0003704 & 0.0000235 & 0.0002480 & 0.0000101 \\
\textcolor{mygray}{\emph{3}} & J100319.37+021447.2 & 150.83073 & 0.00011 & 2.24645 & 0.00013 & 0.0000749 & 0.0000159 & 0.0000785 & 0.0000095 \\
\textcolor{mygray}{\emph{4}} & J100318.49+021000.5 & 150.82705 & 0.00016 & 2.16682 & 0.00021 & 0.0000608 & 0.0000179 & 0.0000582 & 0.0000099 \\
\textcolor{mygray}{\emph{5}} & J100317.73+021606.7 & 150.82388 & 0.00007 & 2.26855 & 0.00008 & 0.0001638 & 0.0000186 & 0.0001419 & 0.0000097 \\
\textcolor{mygray}{\emph{6}} & J100317.32+022016.0 & 150.82219 & 0.00007 & 2.33780 & 0.00009 & 0.0001212 & 0.0000163 & 0.0001200 & 0.0000093 \\
\textcolor{mygray}{\emph{7}} & J100316.70+020457.0 & 150.81962 & 0.00012 & 2.08250 & 0.00024 & 0.0000833 & 0.0000203 & 0.0000683 & 0.0000100 \\
\textcolor{mygray}{\emph{8}} & J100316.52+021207.5 & 150.81886 & 0.00011 & 2.20209 & 0.00014 & 0.0001001 & 0.0000189 & 0.0000876 & 0.0000099 \\
\textcolor{mygray}{\emph{9}} & J100316.52+021810.2 & 150.81884 & 0.00011 & 2.30284 & 0.00016 & 0.0000695 & 0.0000157 & 0.0000708 & 0.0000091 \\
\textcolor{mygray}{\emph{10}} & J100316.10+021700.1 & 150.81712 & 0.00003 & 2.28339 & 0.00003 & 0.0003463 & 0.0000160 & 0.0003498 & 0.0000093 \\
\hline
\end{tabular}
\vspace{2mm}

\begin{tabular}{llllllllll} \hline
 &  $\nu_{\mathrm{eff}}$ & $S^{\mathrm{1.4GHz}}_{\mathrm{int}}$ & $\sigma^{\mathrm{1.4GHz}}_{\mathrm{int}}$ & $S^{\mathrm{1.4GHz}}_{\mathrm{peak}}$ & $\sigma^{\mathrm{1.4GHz}}_{\mathrm{peak}}$ & $\phi_{M}$ & $\sigma_{\phi}^{M}$ & $\phi_{m}$ & $\sigma_{\phi}^{m}$ \\
 & (Hz) & (Jy) & (Jy) & (Jy beam$^{-1}$) & (Jy beam$^{-1}$) & (deg) & (deg) & (deg) & (deg)  \\
 & (9) & (10) & (11) & (12) & (13) & (14) & (15) & (16) & (17) \\ \hline 
\textcolor{mygray}{\emph{1}} & 1304066432 & 0.0000987 & 0.0000275 & 0.0000528 & 0.0000101 & 0.00397 & 0.00087 & 0.00269 & 0.00045 \\
\textcolor{mygray}{\emph{2}} & 1304924800 & 0.0003526 & 0.0000224 & 0.0002361 & 0.0000096 & 0.00364 & 0.00018 & 0.00234 & 0.00008 \\
\textcolor{mygray}{\emph{3}} & 1305698816 & 0.0000713 & 0.0000152 & 0.0000748 & 0.0000090 & 0.00246 & 0.00031 & 0.00222 & 0.00025 \\
\textcolor{mygray}{\emph{4}} & 1306541312 & 0.0000579 & 0.0000170 & 0.0000555 & 0.0000094 & 0.00276 & 0.00053 & 0.00216 & 0.00033 \\
\textcolor{mygray}{\emph{5}} & 1306921600 & 0.0001561 & 0.0000177 & 0.0001352 & 0.0000093 & 0.00261 & 0.00018 & 0.00253 & 0.00017 \\
\textcolor{mygray}{\emph{6}} & 1305002624 & 0.0001154 & 0.0000155 & 0.0001143 & 0.0000089 & 0.00256 & 0.00021 & 0.00225 & 0.00016 \\
\textcolor{mygray}{\emph{7}} & 1306006144 & 0.0000794 & 0.0000193 & 0.0000651 & 0.0000095 & 0.00325 & 0.00058 & 0.00214 & 0.00026 \\
\textcolor{mygray}{\emph{8}} & 1308797184 & 0.0000955 & 0.0000180 & 0.0000835 & 0.0000095 & 0.00279 & 0.00034 & 0.00234 & 0.00024 \\
\textcolor{mygray}{\emph{9}} & 1307299840 & 0.0000662 & 0.0000150 & 0.0000675 & 0.0000087 & 0.00267 & 0.00039 & 0.00210 & 0.00024 \\
\textcolor{mygray}{\emph{10}} & 1308201600 & 0.0003303 & 0.0000152 & 0.0003336 & 0.0000089 & 0.00238 & 0.00006 & 0.00237 & 0.00006 \\
\hline
\end{tabular}
\vspace{2mm}

\begin{tabular}{llllllllllll} \hline
 & $\phi_{\mathrm{PA}}$ & $\sigma_{\phi}^{\mathrm{PA}}$ & $\theta_{M}$ & $\sigma_{\theta}^{M}$ & $\theta_{m}$ & $\sigma_{\theta}^{m}$ & Resolved & $\sigma_{\mathrm{Isl}}$ & ID$_{\mathrm{Gaus}}$ & ID$_{\mathrm{Src}}$ & ID$_{\mathrm{Isl}}$ \\
 & (deg) & (deg) & (deg) & (deg) & (deg) & (deg) &  & (Jy beam$^{-1}$) & &  \\
 & (18) & (19) & (20) & (21) & (22) & (23) & (24) & (25) & (26) & (27) & (28) \\ \hline
\textcolor{mygray}{\emph{1}} & 137.6 & 25.1 & 0.00317 & 0.00109 & 0.00124 & 0.00098 & 0 & 0.0000100 & 0 & 0 & 0 \\
\textcolor{mygray}{\emph{2}} & 179.6 & 4.7 & 0.00274 & 0.00023 & 0.00000 & 0.00000 & 1 & 0.0000097 & 1 & 1 & 1 \\
\textcolor{mygray}{\emph{3}} & 178.8 & 50.2 & 0.00057 & 0.00135 & 0.00000 & 0.00000 & 0 & 0.0000096 & 2 & 3 & 2 \\
\textcolor{mygray}{\emph{4}} & 153.7 & 31.5 & 0.00138 & 0.00106 & 0.00000 & 0.00000 & 0 & 0.0000099 & 3 & 4 & 3 \\
\textcolor{mygray}{\emph{5}} & 178.8 & 89.3 & 0.00104 & 0.00045 & 0.00082 & 0.00053 & 0 & 0.0000094 & 4 & 5 & 4 \\
\textcolor{mygray}{\emph{6}} & 12.7 & 25.8 & 0.00092 & 0.00059 & 0.00000 & 0.00000 & 0 & 0.0000093 & 5 & 6 & 5 \\
\textcolor{mygray}{\emph{7}} & 165.1 & 17.3 & 0.00221 & 0.00085 & 0.00000 & 0.00000 & 0 & 0.0000099 & 6 & 7 & 6 \\
\textcolor{mygray}{\emph{8}} & 20.6 & 28.3 & 0.00144 & 0.00066 & 0.00000 & 0.00000 & 0 & 0.0000097 & 7 & 8 & 7 \\
\textcolor{mygray}{\emph{9}} & 167.3 & 24.1 & 0.00119 & 0.00087 & 0.00000 & 0.00000 & 0 & 0.0000092 & 8 & 9 & 8 \\
\textcolor{mygray}{\emph{10}} & 10.9 & 0.0 & 0.00000 & 0.00000 & 0.00000 & 0.00000 & 0 & 0.0000093 & 9 & 10 & 9 \\
\hline
\end{tabular}
\end{minipage}
\end{table*}

\section{Additional observations in COSMOS}
\label{sec:cosmosappendix}

Table \ref{tab:extra_observations} lists the properties of the 14 additional COSMOS observations (totalling 111 h) that we made use of for the analysis in Section \ref{sec:primarybeams}. The table columns are the same as Table \ref{tab:observations}, and the data were also processed using the methods described in Section \ref{sec:obs_and_proc}.

\begin{table*}
\begin{minipage}{176mm}
\centering
\caption{Details of the additional pointings in the COSMOS field, as plotted in Figure \ref{fig:schematic}, and discussed in Section \ref{sec:primarybeams}.}
\begin{tabular}{lllllllrlll} \hline
Date       & ID         & Field  & RA          & Dec        & Track   & On-source & N$_{\mathrm{chan}}$ & N$_{\mathrm{ant}}$ & Primary & Secondary \\ 
           &            &        &             &            & (h) & (h)   &                     &                    & \\ \hline
2019-07-16&	1563267356&	COSMOS\_1&	09\hhh59\mmm46\sss&	+02\ddd01\dmm44\farcs6&	7	&   6.33&	4096	&59&	J0408-6545&	3C237\\
2019-07-27&	1564215117&	COSMOS\_2&	09\hhh59\mmm46\sss&	+02\ddd22\dmm57\farcs4&	7.95&	6.98&	4096	&61&	J0408-6545&	3C237\\
2019-07-28&	1564301832&	COSMOS\_3&	10\hhh01\mmm11\sss&	+02\ddd01\dmm44\farcs6&	7.96&	6.97&	4096	&60&	J0408-6545&	3C237\\
2019-08-16&	1565939836&	COSMOS\_4&	10\hhh01\mmm11\sss&	+02\ddd22\dmm57\farcs4&	7.99&	6.97&	4096	&58&	J0408-6545&	3C237\\
2020-03-28&	1585413022&	COSMOS\_5&	09\hhh59\mmm04\sss&	+02\ddd12\dmm21\farcs0&	8	&   6.25&	32768	&59&	J0408-6545&	3C237\\
2020-03-29&	1585498873&	COSMOS\_6&	10\hhh01\mmm54\sss&	+02\ddd12\dmm21\farcs0&	8	&   6.25&	32768	&59&	J0408-6545&	3C237\\
2020-03-31&	1585671638&	COSMOS\_7&	10\hhh00\mmm29\sss&	+01\ddd51\dmm08\farcs2&	8	&   6.25&	32768	&60&	J0408-6545&	3C237\\
2020-04-02&	1585844155&	COSMOS\_8&	10\hhh00\mmm29\sss&	+02\ddd33\dmm33\farcs8&	8	&   6.25&	32768	&60&	J0408-6545&	3C237\\
2020-04-30&	1585928757&	COSMOS\_9&	10\hhh01\mmm54\sss&	+02\ddd33\dmm33\farcs8&	8	&   6.25&	32768	&60&	J0408-6545&	3C237\\
2020-04-04&	1586016787&	COSMOS\_10&	09\hhh59\mmm04\sss&	+02\ddd33\dmm33\farcs8&	8.03&	6.25&	32768	&60&	J0408-6545&	3C237\\
2020-04-06&	1586188138&	COSMOS\_11&	09\hhh58\mmm21\sss&	+02\ddd22\dmm57\farcs4&	8	&   6.25&	32768	&59&	J0408-6545&	3C237\\
2020-04-07&	1586274966&	COSMOS\_12&	09\hhh58\mmm21\sss&	+02\ddd01\dmm44\farcs6&	8	&   6.25&	32768	&60&	J0408-6545&	3C237\\
2020-04-12&	1586705155&	COSMOS\_13&	09\hhh59\mmm04\sss&	+01\ddd51\dmm08\farcs2&	8	&   6.25&	32768	&59&	J0408-6545&	3C237\\
2020-04-13&	1586791316&	COSMOS\_14&	10\hhh01\mmm53\sss&	+01\ddd51\dmm08\farcs2&	8	&   6.25&	32768	&60&	J0408-6545&	3C237\\ \hline
\end{tabular}
\label{tab:extra_observations}
\end{minipage}
\end{table*}

\section{Updated mosaic layouts}
\label{sec:mosaics}

\begin{figure*}
\centering
\includegraphics[width=7in]{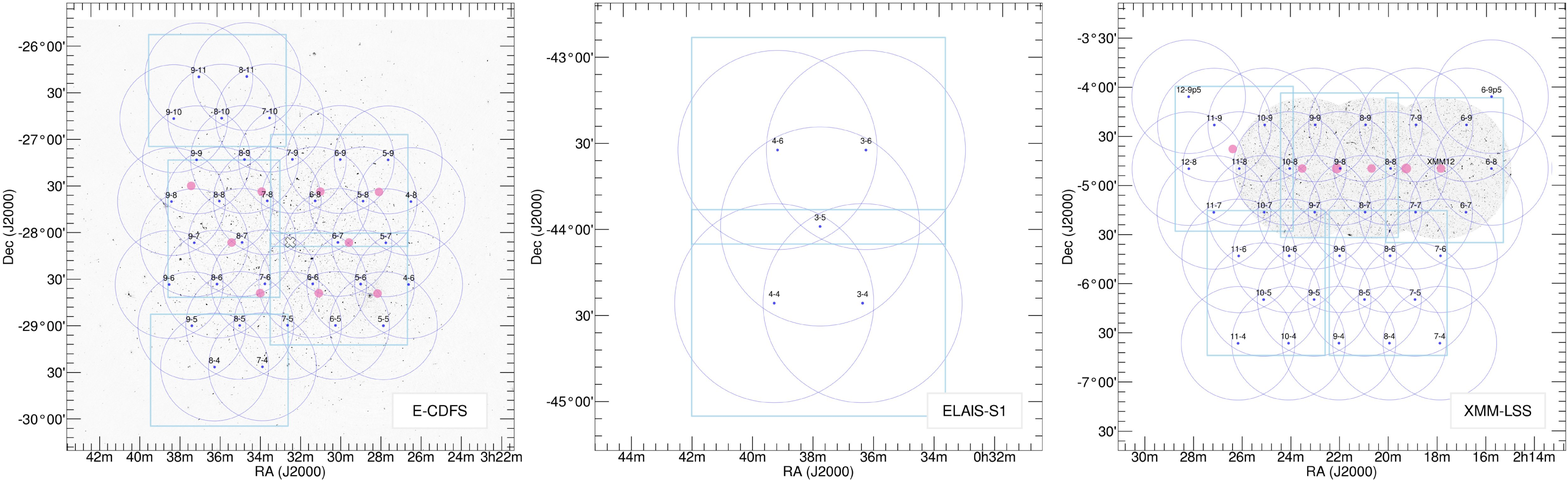}
 \caption{The final pointing layouts (and thus eventual sky coverage) for the MIGHTEE observations of E-CDFS, ELAIS-S1 and XMM-LSS. Please refer to Appendix \ref{sec:mosaics} for further details, and Figure \ref{fig:schematic} and Appendix \ref{sec:cosmosappendix} for the COSMOS layout.}
 \label{fig:full_mosaics}
\end{figure*}

The footprint of the MIGHTEE survey is principally determined by those of existing and planned optical and near-infrared surveys in the four target fields. A secondary consideration is the placement of the pointing centres of the radio observations within these regions. The pointing strategy has been modified since the layout presented in Figure 4 by \citet{jarvis2016}, with the above considerations being joined by our experience with the first batch of MeerKAT observations in each of the four fields. The modifications aim to increase the uniformity of the final survey products, and improve the off-axis polarimetry, mainly by increasing the density of pointings with a corresponding reduction on the time spent on each pointing. 

For XMM-LSS and E-CDFS most of the area is covered by a standard close-packed hexagonal mosaic pattern, which is known to efficiently deliver uniform coverage \citep[e.g.][]{condon1998}. In these two fields each pointing is typically covered by either 2 $\times$ 4 h track, or a single 8 h track (both including overheads), however there will be additional depth in certain areas due to the addition of the initial batch of pointings that were taken as part of the commissioning and Early Science phases. Such additional pointings are marked on the relevant figures. COSMOS and ELAIS-S1 have smaller areas but go deeper due to either having a closer packed mosaic, having 2 $\times$ 8 h tracks on some pointings, or both. The pointing centres for COSMOS are not provided in this appendix as they are already presented in this paper via Table \ref{tab:extra_observations}, and shown in Figure \ref{fig:schematic}.

Figure \ref{fig:full_mosaics} shows the proposed final survey coverage for (left to right) E-CDFS, ELAIS-S1, and XMM-LSS. The blue rectangles show the coverage of the VIDEO \citep{jarvis2013} or VEILS \citep{honig2017} infrared surveys, and are generally representative of the regions that contain the deepest ancillary data from numerous facilities. The labelled pointing centres represent the positions of the final pointing grid, with the circles showing the nominal half-power point of the MeerKAT L-band primary beam at 1284 MHz. The pink markers represent the positions of commissioning or Early Science pointings, the latter featuring in this article, all of which will be used for the final survey products.

Note that for E-CDFS the MIGHTEE pointing grid is snapped to the position of the pointing centre for the ultra-deep LADUMA survey \citep{baker2018}, marked on Figure \ref{fig:full_mosaics} by the central cross marker. The continuum imaging from LADUMA will be included in the final MIGHTEE mosaic, and since most of LADUMA's observations will be conducted with MeerKAT's UHF receivers, there will also be a lower frequency ($\nu_{\mathrm{centre}}$~=~860~MHz) continuum image for that central pointing available for combination with the L-band MIGHTEE data. The background image of the E-CDFS panel on Figure \ref{fig:full_mosaics} shows the continuum image from a single LADUMA UHF track, without any primary beam correction applied, demonstrating the $\sim$2 degree field of view of the UHF imaging (MIGHTEE collaboration, \emph{priv. comm.}). The XMM-LSS Early Science image forms the background to the lower panel of Figure \ref{fig:full_mosaics}, showing the signficant increase in sky area (and cosmological volume) that the full survey will provide. Note that the XMM-LSS pointings are snapped to the original XMMLSS\_12 pointing (see Table \ref{tab:observations}), which has already achieved full depth with the 32,768 channel mode necessary for the spectral line component of MIGHTEE. For planning purposes the pointing centres of the final MIGHTEE survey are provided online as supplementary material.


\bsp	
\label{lastpage}
\end{document}